\documentclass[%
 reprint,
superscriptaddress,
nofootinbib,
 amsmath,amssymb,
pra,
longbibliography
]{revtex4-1}
\usepackage[T1]{fontenc}
\usepackage[utf8]{inputenc}
\usepackage{graphicx}
\usepackage{dcolumn}
\usepackage{bm}
\usepackage{mathrsfs}
\usepackage{amsmath}
\usepackage{float}
\usepackage{soul}
\usepackage{lipsum}
\usepackage{physics}
\usepackage{dsfont}
\usepackage[usenames,dvipsnames]{xcolor}
\usepackage[colorlinks=true,linkcolor=MidnightBlue,citecolor=blue,filecolor=green,urlcolor=PineGreen]{hyperref}
\usepackage{newunicodechar}
\usepackage{tikz}
\usetikzlibrary{decorations.pathmorphing}
\definecolor{electricviolet}{rgb}{0.56, 0.0, 1.0}
\definecolor{amaranth}{rgb}{0.9, 0.17, 0.31}
\newcommand{\T}{\tau}
\newcommand{\ct}{\mathcal{T}}

\def\be{\begin{equation}}
\def\ee{\end{equation}}

\begin{document}

\title{
Stochastic Path Integral Analysis of the \\ Continuously Monitored Quantum Harmonic Oscillator
}
	\author{Tathagata Karmakar}
	\email{tkarmaka@ur.rochester.edu}
	\affiliation{Department of Physics and Astronomy, University of Rochester, Rochester, NY 14627, USA}
	\affiliation{Center for Coherence and Quantum Optics, University of Rochester, Rochester, NY 14627, USA}
		\author{Philippe Lewalle}
		
	\affiliation{Department of Physics and Astronomy, University of Rochester, Rochester, NY 14627, USA}
	\affiliation{Center for Coherence and Quantum Optics, University of Rochester, Rochester, NY 14627, USA}
	\author{Andrew N. Jordan}
	\affiliation{Department of Physics and Astronomy, University of Rochester, Rochester, NY 14627, USA}
	\affiliation{Center for Coherence and Quantum Optics, University of Rochester, Rochester, NY 14627, USA}    
	\affiliation{Institute for Quantum Studies, Chapman University, Orange, CA, 92866, USA}
\date{\today}

\date{\today}

\begin{abstract}
We consider the evolution of a quantum simple harmonic oscillator in a general Gaussian state under simultaneous time-continuous weak position and momentum measurements. We deduce the stochastic evolution equations for position and momentum expectation values and the covariance matrix elements from the system's characteristic function. By generalizing the Chantasri-Dressel-Jordan (CDJ) formalism  (Chantasri et al.~2013 and 2015)  to this continuous variable system, we construct its stochastic Hamiltonian and action. Action extremization gives us the equations for the most-likely readout paths and quantum trajectories.  For steady states of the covariance matrix elements, the analytical solutions for these most-likely paths are obtained. Using the CDJ formalism we calculate final state probability densities exactly starting from any initial state. We also demonstrate the agreement between the optimal path solutions and the averages of simulated clustered stochastic trajectories. Our results provide insights into the time dependence of the mechanical energy of the system during the measurement process, motivating their importance for  quantum measurement engine/refrigerator experiments.
\end{abstract}

\pacs{Valid PACS appear here}
\maketitle
\section{\label{sec:level1}Introduction}
 Methods to implement time--continuous measurements of quantum systems \cite{BookCarmichael, BookWiseman, BookBarchielli, BookJacobs} have emerged in the last three decades; they allow the quantum state to be probed in real time, and are indispensable for feedback control \cite{Zhang2017}. Continuous measurement involves a time sequence of individually weak measurements, in which an observer gently probes the quantum system of interest via its environment, and infers the stochastic evolution of the state as they receive measurement outcomes. In contrast with a projective measurement, time--continuous weak monitoring, leading to diffusive quantum trajectories, causes \emph{gradual} ``collapse'' of the state over time \cite{Jordan2013rev}. Our current work draws together two threads from within this area. On the one hand, there has been considerable recent experimental progress in the continuous monitoring of quantum harmonic oscillators \cite{PhysRevLett.123.163601}. On the other, diffusive quantum trajectories have been analyzed via stochastic path integral methods, which have allowed ``most--likely paths'' (MLPs), following an optimal measurement record, to be computed \cite{chantasri2013action, PhysRevA.92.032125, Areeya_Thesis}. 
These path integral methods, initially developed by Chantasri, Dressel, and Jordan (CDJ), have been applied to theory and experiment across several examples featuring continuously--monitored qubits \cite{chantasri2013action, weber_mapping_2014, PhysRevA.92.032125, Areeya_Thesis, jordan_anatomy_2016, LewalleMultipath, NaghilooCaustic, LewalleChaos, FlorTeach2019}. This formalism has not been applied to the continuously--monitored quantum harmonic oscillator or other continuous variable systems, however. In this work, we build an action based description of the quantum trajectories of general Gaussian oscillator states under time-continuous simultaneous position and momentum measurements, using the CDJ formalism. The more widely studied case of continuous position measurement re-emerges as a special case.\par The harmonic oscillator is ubiquitous across physics leading to analytically soluble models applicable to a wide range of physical systems.   It provides a suitable description of optical degrees of freedom and a wide variety of vibrating elements like cavity mirrors, trapped ions, cantilevers, membranes, and so on. For oscillators, building on the analytical studies of the dynamics of continuous measurement \cite{jacobs2006straightforward,Genoni2016,belenchia_entropy_2020,PhysRevA.60.2700}, quantum state smoothing methods have been explored \cite{PhysRevA.97.042106,laverick_quantum_2019,laverick2020linear,laverick_general_2021,PhysRevA.103.012213}. Hybrid systems, where the quantum state of a mechanical element is probed and readout optically \cite{RevModPhys.86.1391,RevModPhys.82.1155} have received significant attention, too. Experimentally, both position monitoring and control \cite{vanner_towards_2015,PhysRevLett.108.033602,PhysRevLett.117.140401,PhysRevX.7.021008,moller_quantum_2017,PhysRevX.7.011001,PhysRevLett.123.163601, purdy_observation_2013,thompson}, and measurement based feedback cooling of resonators \cite{rossi_measurement-based_2018,wilson_measurement-based_2015,krause2015optical,PhysRevLett.99.017201,kleckner_sub-kelvin_2006,PhysRevLett.99.160801} have been done.

\par Additionally, mechanical resonators have been used to study the quantum phenomena at macroscopic scales, with the majority of the experiments requiring cooling of the resonator as well. 
Recently, experiments have been done to generate entanglement between two mechanical resonators \cite{Riedinger:2018qsr,ockeloen-korppi_stabilized_2018}, between vibrational modes of two diamonds \cite{lee_entangling_2011} and between optical field-mechanical resonator  \cite{riedinger_non-classical_2016,palomaki_entangling_2013}. Complementary to these efforts, superposition states of a membrane resonator \cite{ringbauer_generation_2018} and squeezed states of mechanical resonators  have been generated successfully \cite{lei_quantum_2016, wollman_quantum_2015,PhysRevLett.115.243601,PhysRevX.5.041037, PhysRevLett.112.023601} in optomechanical settings. Furthermore, single phonon generation and second-order phonon correlation calculation for nanomechanical resonators on  Hanbury, Brown, and Twiss setup have been done using optical control \cite{hong_hanbury_2017, cohen_phonon_2015}. \par This paper is organized as follows: in section \ref{stoch_evol}, we describe the stochastic evolution of the system, construct stochastic action and Hamiltonian, and analytically solve the equations of motions for the optimal paths. In section \ref{sec_prob_density} we find the expression for final state probability density using the path integral formalism and in section \ref{energetics} we discuss the energetics of the system. Next, section \ref{OP_traj} explores the connection between stochastic trajectories and optimal paths. Lastly, we present our concluding remarks in section \ref{disc}.
\section{\label{stoch_evol}Stochastic evolution of the system}
\begin{figure}
    \centering
\begin{tikzpicture}[wave/.style={decorate,decoration={snake,pre length = 0.25cm, post length=0.25cm,amplitude=0.8cm,segment length=1.75cm}}]
\node[] at (-.2,-1) {(a)};
\draw[->,red,line width = 0.05cm] (0,-2) -- (2,-2);
\draw[->,red,line width = 0.05cm] (4.5,-2) -- (5.5,-2);
\filldraw[color=blue!60, fill=blue!5, very thick] (1,-2.25) rectangle (1.5,-1.75);
\draw[color=blue!60, very thick] (1.5,-2.25) -- (1,-1.75);
\draw[red,line width = 0.05cm, opacity = 0.5] (1,-2) -- (1.5,-2);
\draw[red,line width = 0.05cm, opacity = 0.5] (1.25,-2) -- (1.25,-2.25);
\draw[->,red,line width = 0.05cm] (1.25,-2.25) -- (1.25,-3.5) -- (5.75,-3.5) -- (5.75,-2.25);
\node[] at (5,-3.2) {\color{red} LO};
\node[] at (5,-1.7) {\color{red} sig.};
\filldraw[color=blue!60, fill=blue!5, very thick] (5.5,-2.25) rectangle (6,-1.75);
\draw[color=blue!60, very thick] (5.5,-2.25) -- (6,-1.75);
\draw[fill = blue!10, draw = blue!10] (2.25,-1) arc (150:210:2) -- (4.25,-3) arc (330:390:2) -- cycle;
\draw[ultra thick, color=blue] (2.25,-1) arc (150:210:2);
\draw[ultra thick, color=blue] (4.25,-3) arc (330:390:2);
\draw[red, line width = 0.05cm, opacity = 0.5] (5.75,-2.25) -- (5.75,-1.75);
\draw[red, line width = 0.05cm, opacity = 0.5] (5.75,-1.75) -- (5.75,-1);
\draw[red, line width = 0.05cm, opacity = 0.5] (6,-2) -- (6.75,-2);
\draw[red, line width = 0.05cm, opacity = 0.5] (5.5,-2) -- (6,-2);
\draw[black, opacity = 0.2, line width = 0.05cm] (3.25,-1) arc (150:210:2);
\draw[black, opacity = 0.35, line width = 0.05cm] (3.25,-1) -- (3.25,-3);
\draw[black, opacity = 0.5, line width = 0.05cm] (3.25,-3) arc (330:390:2);
\draw[black, dashed, ultra thick] (6.75,-2) -- (7,-2);
\draw[black, dashed, ultra thick] (5.75,-1) -- (5.75,-.75);
\draw[black, dashed, ultra thick] (5.75,-1) -- (5.75,-.75);
\draw[black, dashed, ultra thick] (5.75,-.75) -- (6.85,-.75);
\draw[black, dashed, ultra thick] (7,-2) -- (7,-.90);
\draw[thick, black] (7,-.75) circle (0.15);
\node[] at (7,-.75) {--};
\draw[red, wave, opacity = 0.5, line width = 0.05cm] (2,-2) -- (4.5,-2);
\node[] at (3.25,-2) {\color{black!50!red} \Large $\otimes$};
\draw[fill = black,color= black,rounded corners = 0.05cm] (6.5,-1.75) arc[radius = 0.25, start angle = 90, end angle = -90] -- cycle;
\draw[fill = black,color= black,rounded corners = 0.05cm] (5.5,-1.25) arc[radius = 0.25, start angle = 180, end angle = 0] -- cycle; 
\end{tikzpicture}

\begin{tikzpicture}[wave/.style={decorate,decoration={snake,pre length = 0.25cm, post length=0.25cm,amplitude=0.8cm,segment length=1.75cm}}]

\node[] at (0,0) {\includegraphics[width=0.5\textwidth,trim = {30 300 30 200}, clip]{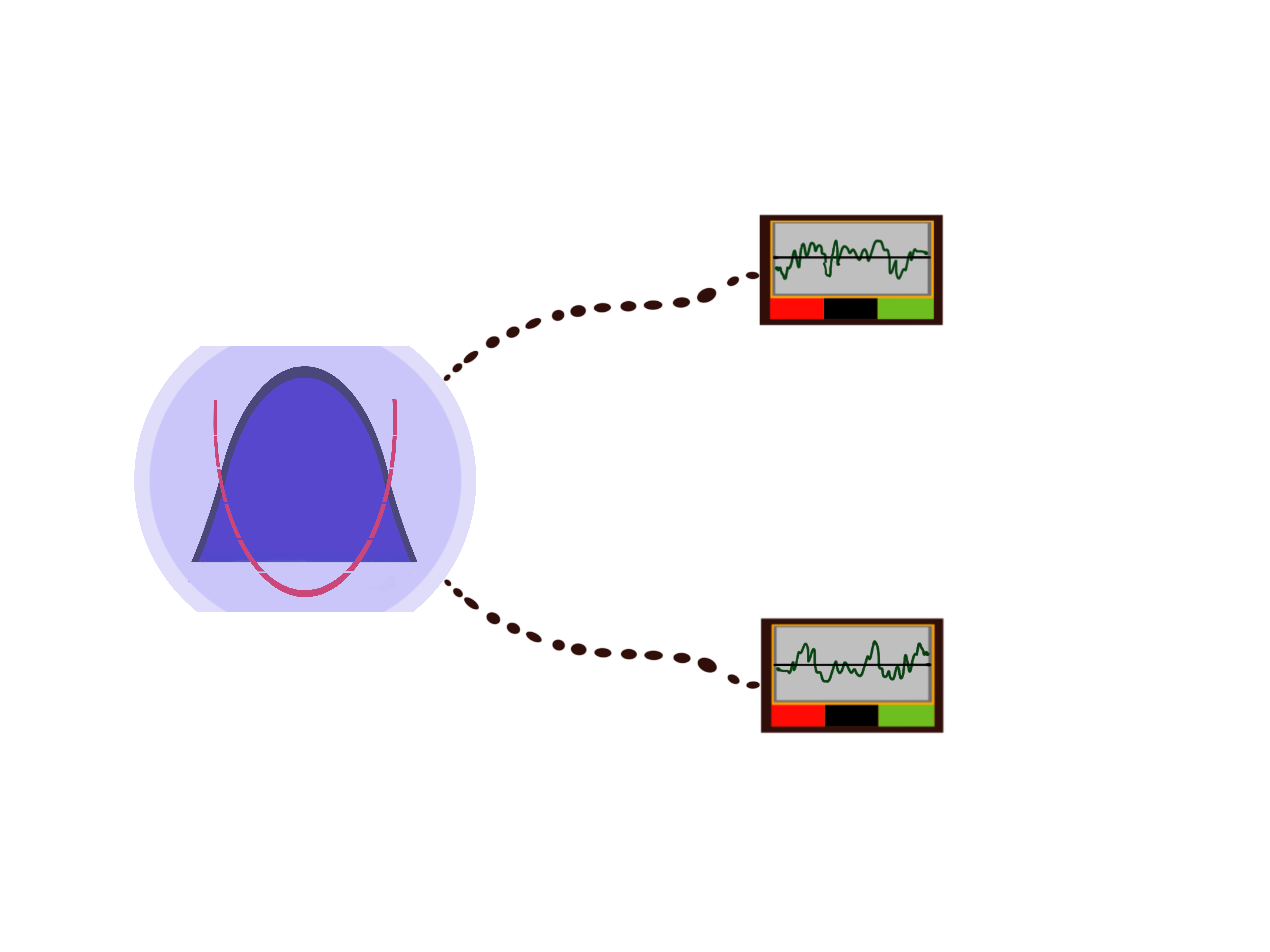}};
\node[] at (-3.78,1.9) {\color{black} (b)};
\node[] at (-2.4,-1.5) {\color{black} Harmonic};
\node[] at (-2.4,-1.8) {\color{black} Oscillator};

\node[] at (2.6,1.3) {\color{black} $r_1$};
\node[] at (-1,1) {\color{black} $X$};
\node[] at (2.6,-1.65) {\color{black}
$r_2$};
\node[] at (-1,-1.5) {\color{black} $P$};
\end{tikzpicture}
\vspace*{-.3cm}
    \caption{We sketch an example system of interest, a mechanical oscillator, being probed using an optical field.   Figure (a) shows the case where the position of the oscillator is continuously monitored by performing homodyne detection on the transmitted light. Figure (b) shows the schematic of the setup for simultaneous position and momentum measurements. In this case, the harmonic oscillator under consideration is coupled to two measurement devices which independently readout position and momentum of the oscillator simultaneously and continuously. A similar measurement scheme has been implemented successfully for qubits \cite{Leigh2016}.}
    \label{fig-device}
\end{figure}
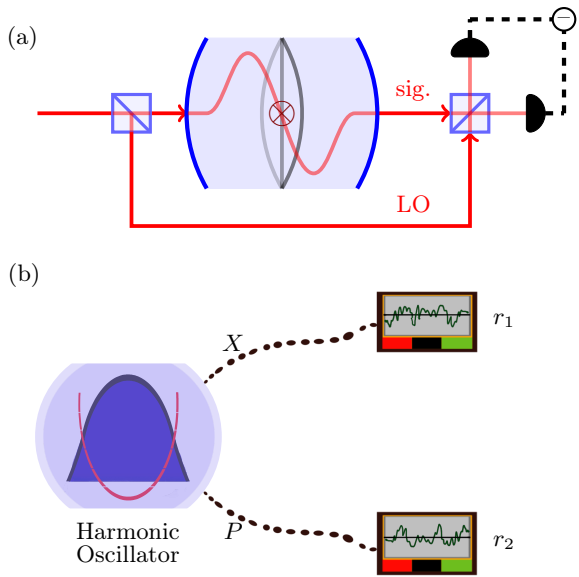

Our system of interest is a harmonic oscillator  which is being time-continuously monitored via two detectors. An example, shown in Fig.~\ref{fig-device}, is a simple harmonic oscillator and its position and momentum are simultaneously and continuously monitored by an optical measurement set-up.
\subsection{\label{state_evol}State Description and Evolution} 
The closed quantum harmonic oscillator exhibits unitary evolution due to its Hamiltonian.
\begin{equation}
    \hat{H}=\frac{1}{2m}\hat{p}^2+\frac{1}{2}m\omega^2\hat{x}^2.
    \label{hamiltonian}
\end{equation}
We define the dimensionless position and momentum observables as $\hat{X}=\sqrt{\frac{m\omega}{\hbar}}\hat{x}$ and $\hat{P}=\frac{\hat{p}}{\sqrt{\hbar m\omega}}$. We also define  $\tau=\omega t$ and $\hat{H}^\prime=\frac{\hat{H}}{\hbar\omega}$. From now on, we will refer to these four dimensionless quantities as position, momentum, time and Hamiltonian (or energy) respectively.\par
For the experiments of the type in Fig.~\ref{fig-device}, we consider weak continuous measurements of a Gaussian type. The Kraus operators \cite{BookKraus} representing variable strength position or momentum measurements are given, respectively, by
\begin{subequations}
\begin{equation}
    \hat{M}_{X}(r_1)=\Big(\frac{\delta \tau}{2 \pi \tau_1}\Big)^\frac{1}{4}\exp\left[-\frac{\delta \tau}{4\tau_1}(r_1\mathds{1}-\hat{X})^2\right]\label{kraus_x}
\end{equation}
and
\begin{equation}
    \hat{M}_{P}(r_2)=\Big(\frac{\delta \tau}{2 \pi \tau_2}\Big)^\frac{1}{4}\exp\left[-\frac{\delta \T}{4\tau_2}(r_2\mathds{1}-\hat{P})^2\right],
    \label{kraus_p}
\end{equation}
\end{subequations}
with $r_1$ and $r_2$ as the readouts of the two measurements. Here $\delta \T$ is the duration of the measurement. We only consider weak measurements (long collapse timescales) i.e. $\delta \T \ll\tau_1,$ $\tau_2$.\par 
If measurements are made over a time interval from $\tau$ to $\T+\delta \T$, 
the state (density matrix $\hat{\rho}$) can then be updated via \cite{BookCarmichael, BookWiseman, BookBarchielli, BookJacobs} $\hat{\rho}(\T+\delta \T)=$
\begin{equation}
    \frac{e^{-i\hat{H}^\prime\delta \T}\hat{M}_{X}(r_1) \hat{M}_{P}(r_2)\hat{\rho}(\tau)\hat{M}_{P}^{\dagger}(r_2)\hat{M}_{X}^{\dagger}(r_1)e^{i\hat{H}^\prime\delta \T}}{\text{Tr}[\hat{M}_{X}(r_1) \hat{M}_{P}(r_2)\hat{\rho}(\T)\hat{M}_{P}^{\dagger}(r_2)\hat{M}_{X}^{\dagger}(r_1)]},
    \label{rho_evol_xp}
\end{equation}
which includes the update conditioned on the outcomes of simultaneous weak $\hat{X}$ and $\hat{P}$ measurements, as well as the unitary evolution due to the bare Hamiltonian \eqref{hamiltonian}. The choice of order of operations above is arbitrary, as the commutators between the different operators only play a role to $O(\delta \T^2)$, 
and can be neglected in the limit of weak continuous measurement. 
Although the Heisenberg uncertainty principle forbids \emph{joint projective} measurement of complimentary observables, simultaneous weak (non-projective) measurement can be done as described by Jordan and  B\"uttiker for qubits \cite{Jordan2005}. Notably, the protocol by Arthurs and Kelly for harmonic oscillators provides a way to achieve the minimum uncertainty limit for such simultaneous measurements \cite{AK_1965,Ochoa2018}.
Physical implementations of this process on optical degrees of freedom are typically achieved via heterodyne detection \cite{shapiro_quantum_2009}, or quantum--limited phase--preserving amplification \cite{Caves2012} (either of which can be associated with projectors in the coherent state basis). 
Further ideas about continuous monitoring of non--commuting observables have been developed in the literature concerning qubit measurements as well (see e.g.~\cite{ Ruskov2010, Leigh2016, LewalleMultipath, LewalleChaos, Atalaya2018, Atalaya2018_2, Atalaya2019}).

\par Our analysis is restricted to general Gaussian states (see Appendix \ref{appA0}) of the oscillator. Common states, including coherent states, and measurements are both Gaussian, making  such restriction a reasonable choice \cite{Genoni2016, jacobs2006straightforward, zhang_prediction_2017, wang2007quantum}. We will see below that both unitary and measurement dynamics of type \eqref{rho_evol_xp} preserves the Gaussian form of the states. Such states' evolution can be expressed in terms of the expectation values of position and momentum, their variances, and their covariance. The characteristic function at time $\T$, $\chi_\T(\pmb{\xi})$ (see Appendix \ref{appA0}), is by definition a Gaussian in $\pmb{\xi}=(\xi_1,~\xi_2)^\top$. At time $\T+\delta \T$ the characteristic function will be (from \eqref{char_fn}), 
\begin{equation}
    \chi_{\T+\delta \T}(\pmb{\xi})=\text{Tr}\big(\hat{\rho}(\T+\delta \T)e^{i\pmb{\xi}^\top\cdot\pmb{\hat{\mathcal{R}}}}\big),
    \label{char_deltat}
\end{equation}
with $\pmb{\hat{\mathcal{R}}}=(\hat{X},~\hat{P})^\top$. It can be shown that to $O(\delta \T)$ we have
\begin{subequations}
\begin{equation}
    \chi_{\T+\delta \T}(\pmb{\xi})=\chi_\T(\pmb{\xi})e^{ \kappa\delta \T},
    \label{char_evo}
\end{equation}
with $\kappa$ given as
\begin{equation}
    \begin{split}
    &\frac{1}{2}(q_3\xi_1+q_4\xi_2)\xi_2-\frac{1}{2}(q_4 \xi_1+q_5\xi_2)\xi_1 -\frac{1}{8\tau_2}\xi_1^2-\frac{1}{8\tau_1}\xi_2^2\\& +\frac{1}{8\tau_1}(q_3\xi_1+q_4\xi_2)^2+\frac{1}{8\tau_2}(q_4\xi_1+q_5\xi_2)^2 -i q_1\xi_2+i\xi_1 q_2\\&+\frac{i}{8\tau_1}(q_3\xi_1+q_4\xi_2)(r_1-q_1)+\frac{i}{8\tau_2}(q_4\xi_1+q_5\xi_2)(r_2-q_2).
\end{split}
\label{kappa}
\end{equation}
\end{subequations}
The variables $q_1$ to $q_5$ are defined as the scaled cumulants 
\begin{equation}
    \begin{split}
        &q_1=\expval{\hat{X}},\:q_2=\expval{\hat{P}},\:q_4=\expval{\hat{X}\hat{P}+\hat{P}\hat{X}}-2\expval{\hat{X}}\expval{\hat{P}},\\
        & q_3=2\left(\expval{\hat{X}^2}-\expval{\hat{X}}^2\right),\:\:\:\&\:\:\:q_5=2\left(\expval{\hat{P}^2}-\expval{\hat{P}}^2\right).
    \end{split}
    \label{defns}
\end{equation}
 The form of \eqref{kappa} confirms the preservation of Gaussianity of the state \cite{Genoni2016, zhang_prediction_2017, laverick_quantum_2019}. 
Equivalently, we can characterize the state evolution in terms of the variables defined in \eqref{defns}. From \eqref{char_gaussian}, \eqref{char_evo} and \eqref{kappa}, the stochastic equations in $q_1$ and $q_2$ can be written in the Langevin form
\be \label{q_stoch}
\dot{\pmb{q}} = \boldsymbol{\mathsf{\Omega}}\cdot\pmb{q} + \boldsymbol{\mathsf{b}}\cdot \frac{d\mathbf{W}}{d\T} = \boldsymbol{\mathsf{\Omega}} \cdot \pmb{q} + \boldsymbol{\mathsf{b}}\cdot(\boldsymbol{\mathsf{R}} - \boldsymbol{\mathsf{X}})=\pmb{\mathcal{F}}(\pmb{q},\pmb{r}),
\ee 
where $\pmb{q}=(q_1,q_2)^\top$; $\boldsymbol{\mathsf{\Omega}}$ and $\boldsymbol{\mathsf{b}}$ are the matrices
\begin{equation}
    \boldsymbol{\mathsf{\Omega}} =\begin{pmatrix}0 & 1 \\-1 &0\end{pmatrix},\:\:\:\: \boldsymbol{\mathsf{b}} =\begin{pmatrix}\frac{q_3}{2\sqrt{\tau_1}} &\frac{q_4}{2\sqrt{\tau_2}} \\[6pt]\frac{q_4}{2\sqrt{\tau_1}} &\frac{q_5}{2\sqrt{\tau_2}}\end{pmatrix}.
    \label{Omega}
\end{equation}$d\mathbf{W}$ is the vector of Wiener noises $dW_1$ and $dW_2$. They are related to the measurement readouts by 
\begin{subequations}
\begin{equation}
        r_1=\expval{\hat{X}}+\sqrt{\tau_1}\frac{dW_1}{d\T},\:\:\:\:
        r_2=\expval{\hat{P}}+\sqrt{\tau_2}\frac{dW_2}{d\T}, 
        \label{weiner}
\end{equation} which may each be interpreted as a sum of signal and noise terms. Here $r_1$ measures the position of the oscillator, while $r_2$ measures its momentum. For notational convenience we have defined $ \boldsymbol{\mathsf{R}} = \boldsymbol{\mathsf{X}}+d\mathbf{W}/dt $ with \begin{equation}
    \boldsymbol{\mathsf{R}} = \left(\begin{array}{c} r_1/\sqrt{\tau_1} \\ r_2/\sqrt{\tau_2} \end{array} \right), \quad \boldsymbol{\mathsf{X}} = \left(\begin{array}{c} \expval{\hat{X}}/\sqrt{\tau_1} \\ \expval{\hat{P}}/\sqrt{\tau_2} \end{array}\right).
\end{equation}
\label{noise}%
\end{subequations} The diffusion tensor $\boldsymbol{\mathsf{b}}$  depends on $q_3$, $q_4$, and $q_5$. It will later be useful to also shorthand $\boldsymbol{\mathsf{B}} \equiv \boldsymbol{\mathsf{b}}\,\boldsymbol{\mathsf{b}}^\top$. $\boldsymbol{\mathsf{B}}$ is a square, positive, symmetric matrix by construction.
The covariance matrix elements evolve according to the deterministic equations
\begin{subequations}%
\begin{align}
\dot{q}_3 &=2q_4-\frac{1}{2\tau_1}q_{3}^2-\frac{1}{2\tau_2}q_{4}^2+\frac{1}{2\tau_2},
    \label{q3evol}\\
    \dot{q}_4 &=q_5-q_3-\frac{1}{2\tau_1}q_3 q_4-\frac{1}{2\tau_2}q_4 q_5,
    \label{q4evol}\\
    \dot{q}_5 &=-2q_4-\frac{1}{2\tau_1}q_{4}^2-\frac{1}{2\tau_2}q_{5}^2+\frac{1}{2\tau_1},
    \label{q5evol}
\end{align}
\label{q_deter}%
\end{subequations}consistent with \cite{Genoni2016, jacobs2006straightforward}. \par
These equations of motion are equivalent to those obtained from a stochastic master equation, which can be derived from \eqref{rho_evol_xp}, and whose It\^{o} form reads \cite{jacobs2006straightforward}
\begin{equation}
\begin{split}
    & d\rho
    =-i[\hat{H}^\prime,\hat{\rho}]d\T+ \frac{1}{4\tau_1}\left[ \hat{X}\hat{\rho}\hat{X}-\frac{1}{2}(\hat{X}^2\hat{\rho}+\hat{\rho}\hat{X}^2)\right]d\T+\\&\frac{1}{4\tau_2}\left[ \hat{P}\hat{\rho}\hat{P}-\frac{1}{2}(\hat{P}^2\hat{\rho}+\hat{\rho}\hat{P}^2)\right]d\T+\\&\frac{1}{\sqrt{4\tau_1}}[ \hat{X}-\expval{\hat{X}}, \hat{\rho}]_+dW_1+\frac{1}{\sqrt{4\tau_2}}[ \hat{P}-\expval{\hat{P}}, \hat{\rho}]_+dW_2.
\end{split}
    \label{master}
\end{equation}
\begin{figure}
	\includegraphics[width=0.90\linewidth,trim = {0 0 0 20}, clip]{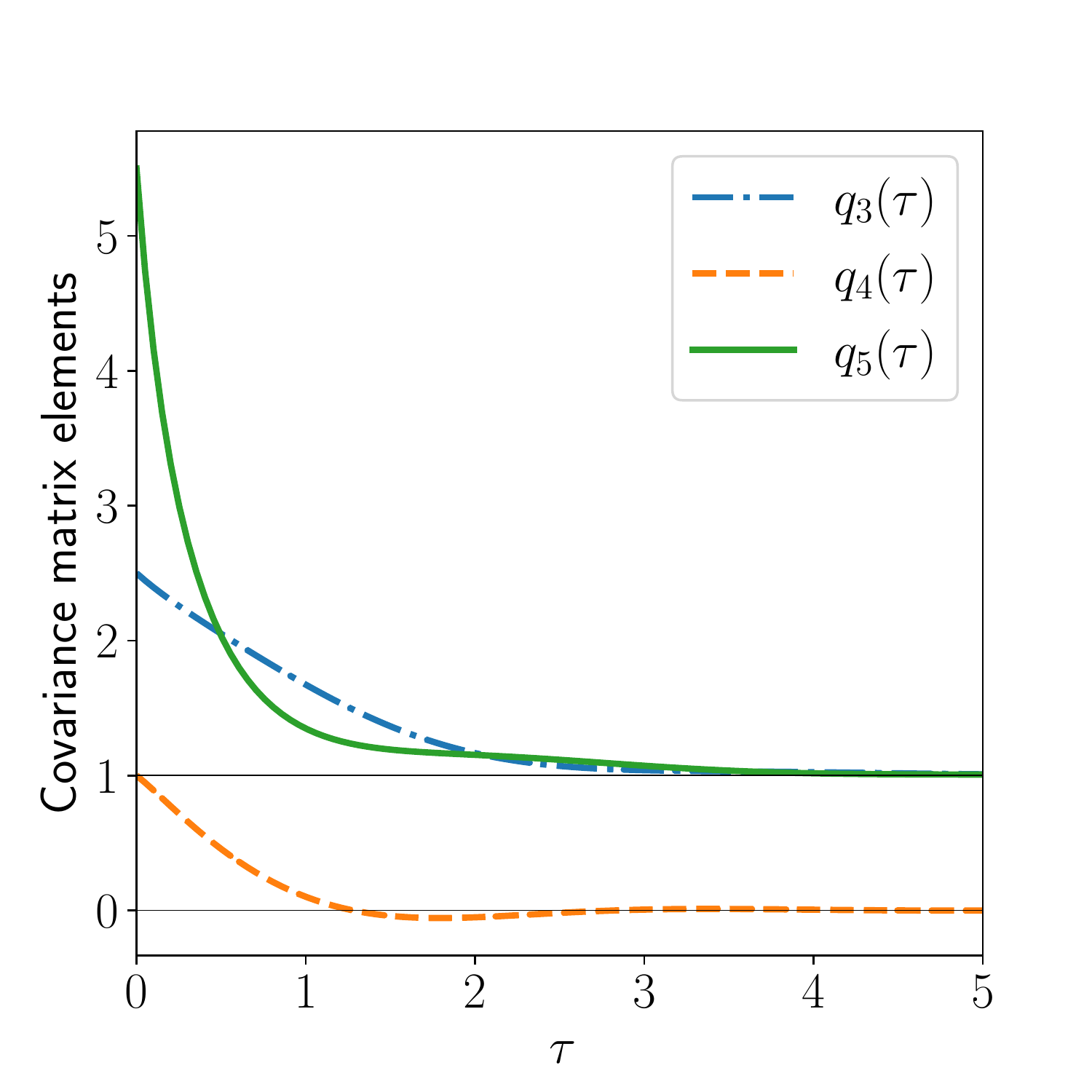}
	\caption{Time evolution of covariance matrix elements for equal measurement strength ($\tau_1=\tau_2=1$). The blue dashdotted line shows $q_3$, the orange dashed line shows $q_4$, and the green solid line corresponds to $q_5$ defined in \eqref{defns}. The initial values are $q_3(0) = 2.5$, $q_4(0) = 1$, and $q_5(0) = 5.5$. The covariance matrix elements tend to the fixed points \eqref{fpoints_special} for $\T\to\infty$. As the determinant goes to 1, this evolution additionally represents purification of the state by continuous measurement.}
	\label{covplot77}%
\end{figure}Here $[...,...]_+$ denotes the anticommutator of two observables. It is noteworthy that no distinction needs to be made between the It\^{o} and Stratonovich forms of \eqref{q_stoch} and \eqref{q_deter} for these Gaussian state equations.
\par Unlike the quadrature variables $q_1$ and $q_2$, the covariance matrix elements $q_{3}$, $q_4$, and $q_5$, evolve deterministically, as they depend neither directly nor indirectly on the measurement record(s). 
Therefore, their evolution is governed by ordinary, rather than stochastic differential equations. 
Typical behavior of these covariance matrix elements is shown in Fig. \ref{covplot77}. 
We can see that for large $\T$, all of them tend to the fixed points values of \eqref{q_deter} (proof of this statement is shown in Appendix \ref{appA1}). For arbitrary measurement strength, the fixed points to which they settle are
\begin{subequations}
\begin{equation}
    \begin{split}
        \tilde{q}_{4}&=\frac{1}{2\gamma_2}\left(\sqrt{\frac{\gamma_1^2}{\tau_2^2}+4\gamma_2^2}-\frac{\gamma_1}{\tau_2}\right),\\
        \tilde{q}_{3}&=\sqrt{\frac{4\tau_1\tilde{q}_4}{\gamma_2}\left(1+\frac{1}{4\tau_2^2}\right)},\\
        \tilde{q}_{5}&=\sqrt{(1+4\tau_1^2)\frac{\tilde{q}_{4}}{\gamma_2\,\tau_1}}.
    \end{split}
    \label{fpoints_xp}
\end{equation}
We have used the definitions $\gamma_1=1+4\tau_1\,\tau_2$ and $\gamma_2=1-\tau_1/\tau_2$. For $\tau_1=\tau_2$ (equal measurement strength) the fixed points are 
\begin{equation}
    \tilde{q}_{3}=1,\: \tilde{q}_{4} = 0,\: \tilde{q}_{5}=1.
    \label{fpoints_special}
\end{equation}
\end{subequations}
The oscillator as a whole tends to a pure state as $q_3 q_5-q_4^2\to 1$ (see Appendices  \ref{appA0} and \ref{appA1}). For the case of equal measurement strength emphasized in this work, the covariance matrix converges to the identity matrix and we get a coherent state at $\T\to\infty$ \cite{wang2007quantum}. In what follows, we take the covariance matrix elements to have these fixed values for simplicity and ignore the time evolution of $q_3$, $q_4$, and $q_5$, that is, we start our time after the covariance matrix reaches its steady state. 
\begin{figure}
	\includegraphics[width=0.9\linewidth,trim = {0 0 0 30}, clip]{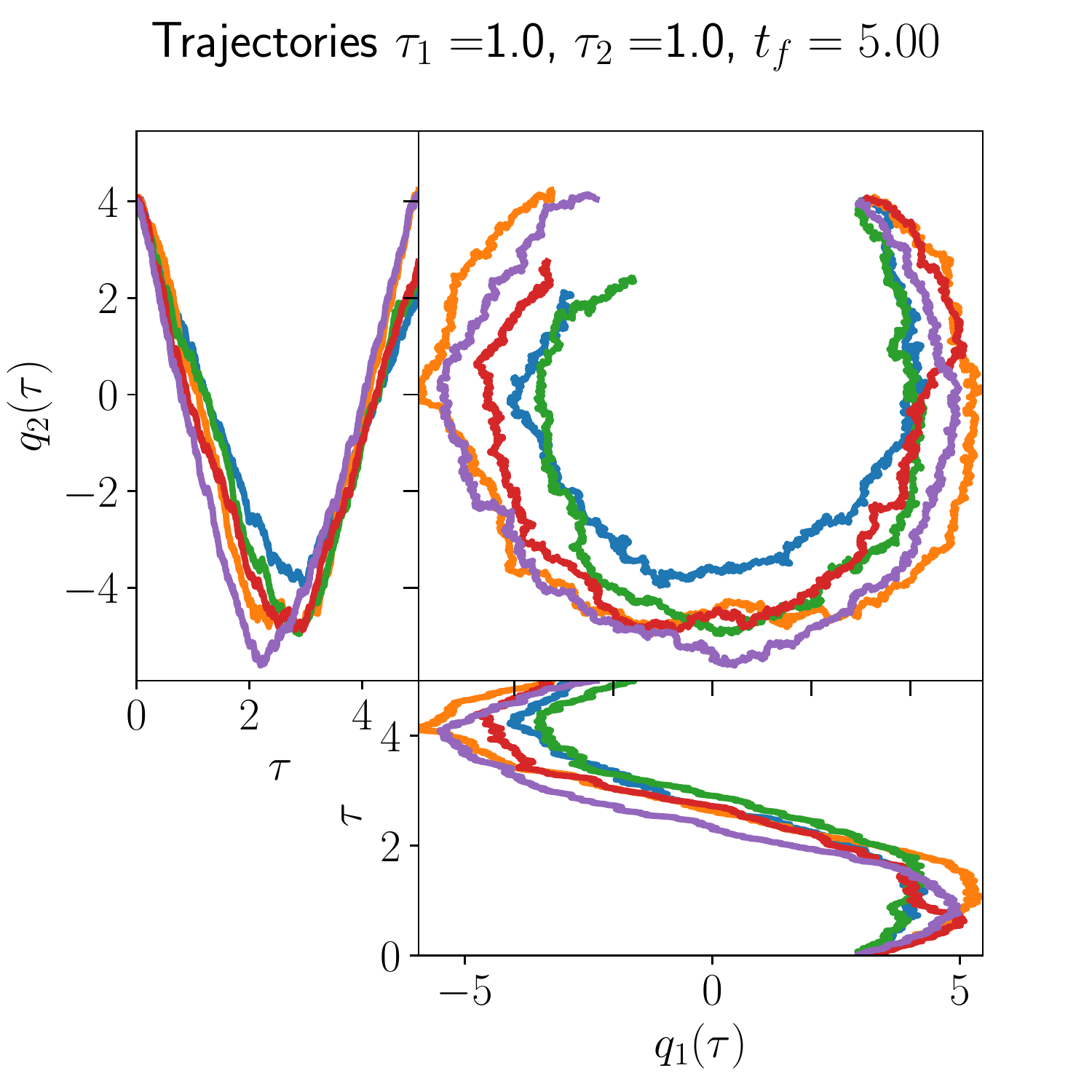}
	\caption{Sample stochastic trajectories till final time  $\T_f=5.0$ for $\tau_1=\tau_2=1$ given initial conditions $q_1(0)=3$  and $q_2(0)=4$. The top left panel shows $q_2$ vs.~time, the bottom right panel shows time vs.~$q_1$, and the top right panel shows $q_2$ vs.~$q_1$. The spiral patterns in the trajectories arise from diffusion about the oscillatory unitary dynamics.}
	\label{init_traj_11}
\end{figure}
\par We now begin investigating the stochastic evolution of the quadratures. We use the Euler method to integrate  \eqref{q_stoch}. The global truncation error is proportional to $\delta \T$. Fig.~\ref{init_traj_11} shows
 sample stochastic trajectories, each generated from the same initial condition, where we see diffusion about the phase space ellipses given by the unitary evolution.
\subsection{\label{stoch_H}Stochastic action and Hamiltonian}
We briefly recap the CDJ (Chantasri, Dressel, Jordan) stochastic path integral (SPI) formalism \cite{chantasri2013action,PhysRevA.92.032125,Areeya_Thesis} and adapt it to the case of interest. 
The state at time $\T_k$ is given by $\rho_k$ (which in turn is completely characterized by $\pmb{q}_k$), and the initial and final $\pmb{q}$ are given by $\pmb{q}_i$ and $\pmb{q}_f$. The joint probability density of having the readouts \{$r_{1k},\:r_{2k}$\} and states $\pmb{q}_k$ is given by
\begin{equation}
    \mathcal{P}=\delta^2(\pmb{q}_0-\pmb{q}_i)\delta^2(\pmb{q}_n-\pmb{q}_f)\prod_{k=0}^{n-1} P(\pmb{q}_{k+1},\pmb{r}_{k}|\pmb{q}_k).
    \label{p_density_xp}
\end{equation}
We can write the conditional probability on the right hand side of \eqref{p_density_xp} as $P(\pmb{q}_{k+1},\pmb{r}_{k}|\pmb{q}_k)=P(\pmb{q}_{k+1}|\pmb{q}_k,\pmb{r}_{k})P(\pmb{r}_{k}|\pmb{q}_k)$. For the purposes of writing the stochastic path integral, we only need to focus on the variables which are in fact stochastic. The stochasticity of the equations is entirely contained in the readouts drawn from the probability density
\begin{equation}
\begin{split}
     P(\pmb{r}_{k}|\pmb{q}_k)&=P(\pmb{r}_{k}|\rho_k)=\\&\text{Tr}[\hat{M}_{X}(r_{1k}) \hat{M}_{P}(r_{2k})\hat{\rho}(\T_k)\hat{M}_{P}^{\dagger}(r_{1k})\hat{M}_{X}^{\dagger}(r_{2k})],
\end{split}
    \label{readout-prob_xp}
\end{equation}
and the coordinates $q_1$ and $q_2$ which evolve as a function of those readouts. Those quadrature coordinates evolve deterministically when we condition on \emph{the values of the stochastic readouts $r_1$ and $r_2$}, and are thereby described by 
$P(\pmb{q}_{k+1}|\pmb{q}_k,\pmb{r}_{k})=\delta^2(\pmb{q}_{k+1}-\pmb{q}_{k}-\delta \T\,\pmb{\mathcal{F}}(\pmb{q}_k,\pmb{r}_{k}))$. The remaining three coordinates (describing the covariance matrix) evolve deterministically (independently of $r_1$, $r_2$, $q_1$, and $q_2$) and can therefore largely be ignored in constructing the path integral; they appear as time--dependent driving terms if allowed to evolve, or merely as constants if they are at their steady state.  
\par Each of the delta functions associated with $P(\pmb{q}_{k+1}|\pmb{q}_k,\pmb{r}_k)$, as well as the boundary conditions in \eqref{p_density_xp}, can be expressed as Fourier integrals of the form $\delta(q)=\tfrac{1}{2\pi i}\int_{-i\infty}^{i\infty}\exp{-pq}dp$. With this in mind can write $P(\pmb{r}_{k}|\pmb{q}_k)\approx \exp[C+\delta \T\,\mathcal{G}(\pmb{q}_k,\pmb{r}_{k})+O(\delta \T^2)]$ where $C = \log(\delta \T/2\pi \sqrt{\tau_1\,\tau_2})$ and
\begin{equation}
\begin{split}
    \mathcal{G}(\pmb{q},\pmb{r})&= -\frac{(r_1 - q_1)^2}{2\tau_1} - \frac{(r_2 -  q_2)^2}{2\tau_2} - \frac{q_3}{4\tau_1} - \frac{q_5}{4\tau_2}, 
\end{split}
    \label{function_f_xp}
\end{equation}
where the first two terms indicate that the noise on each measurement is Gaussian (i.e.~noise about the means $q_1$ or $q_2$ is characterized by a variance $\tau_1/\delta \T$ or $\tau_2/\delta \T$, respectively), consistent with \eqref{noise}. 
The last two terms $\mathsf{g}_1 \equiv q_3/4\tau_1 = \text{var}(\hat{X})/2\tau_1$ and $\mathsf{g}_2 \equiv q_5/4\tau_2 = \text{var}(\hat{P})/2\tau_2$ can be interpreted as a measure of the local rate of information gain expected from each measurement\footnote{A larger variance in e.g.~$\hat{X}$ indicates a greater uncertainty about the outcome of the $\hat{X}$ measurement. As a general principle, the more uncertainty there is about the outcome of a measurement, the more information one gains by performing that measurement. Thus we identify $\mathsf{g}_1$ and $\mathsf{g}_2$ as specifying the rate of an observer's information gain by making $\hat{X}$ and $\hat{P}$ measurements with strengths $\tau_1$ and $\tau_2$, respectively. See \cite{OM-CDJ-SPI} for related discussion.}$^,$\footnote{Note also that a straightforward application of the Heisenberg uncertainty principle $q_3\, q_5 \geq 1$ indicates that $\mathsf{g}_1 \,\mathsf{g}_2 \geq (16\,\tau_1\,\tau_2)^{-1}$. } \cite{BookBarchielli}.
Using Fourier forms of the $\delta$--functions (thereby introducing a pair of variables $p_1$ and $p_2$ conjugate to $q_1$ and $q_2$) we may now express the RHS of \eqref{p_density_xp} as a stochastic path integral; taking a time--continuum limit \cite{chantasri2013action,PhysRevA.92.032125}  leads to 
\begin{equation}
    \mathcal{P}=\int\mathcal{D}\pmb{p}\:e^\mathcal{S}=\int\mathcal{D} \pmb{p}\: \exp\left[\int_{0}^{\T_f}d\T(-\pmb{p}\cdot\pmb{\dot{q}}+\mathcal{H}(\pmb{p},\pmb{q},\pmb{r}))\right],
    \label{action_integral}
\end{equation}
where a stochastic action $\mathcal{S}$ has been defined in terms of a stochastic Hamiltonian
\begin{equation}
\begin{split}
    \mathcal{H}
    =& \pmb{p}^\top\cdot\pmb{\mathcal{F}}(\pmb{q},\pmb{r})+\mathcal{G}(\pmb{q},\pmb{r})\\ 
    =& \pmb{p}^\top\cdot\left(\boldsymbol{\mathsf{\Omega}}\cdot\pmb{q}+\boldsymbol{\mathsf{b}}\cdot[\boldsymbol{\mathsf{R}} - \boldsymbol{\mathsf{X}}]\right)-\tfrac{1}{2}(\boldsymbol{\mathsf{R}}-\boldsymbol{\mathsf{X}})^2 - \mathsf{g}.
\end{split}
    \label{stoch_h_XP}
\end{equation}
Here $\pmb{p}=(p_1,p_2)^\top$.
We have shorthanded $\mathsf{g} = \mathsf{g}_1 + \mathsf{g}_2 = q_3/4\tau_1 + q_5/4\tau_2$. We stress that $\pmb{p}$ and $\mathcal{H}$ are unrelated to the mechanical momentum and Hamiltonian. 
\subsection{\label{eq_OP}Equation of Motion for Optimal Path} The CDJ formalism guides us to the action $\mathcal{S}$. 
Given initial and final states, the variational solutions $\delta \mathcal{S} = 0$ are trajectories following extremal--probability measurement records that connect the given boundary conditions \eqref{p_density_xp}. The equations for these trajectories can be summarized as the three equations $\partial_{\boldsymbol{\mathsf{R}}}\mathcal{H}=0$, $\pmb{\dot{q}}=\partial_{\pmb{p}}\mathcal{H}$, and
    $\pmb{\dot{p}}=-\partial_{\pmb{q}}\mathcal{H}$. We will refer to the corresponding solutions as optimal paths (OPs), or as most--likely paths (MLPs) when they maximize the record probability \cite{chantasri2013action, LewalleMultipath}.\par
The readouts corresponding to the optimal paths (also called optimal readouts) can be calculated from $\partial_{\boldsymbol{\mathsf{R}}}\mathcal{H}=0$, giving 
\begin{equation}
\begin{split}
    \boldsymbol{\mathsf{R}^\star} =\boldsymbol{\mathsf{X}} + \boldsymbol{\mathsf{b}}^\top\cdot \pmb{p},
\end{split}
    \label{OPreadout-val_XP}
\end{equation}
where $\boldsymbol{\mathsf{X}}$ is the signal and $\boldsymbol{\mathsf{b}}^\top\cdot \pmb{p}$ is the optimal noise.\par In the event that the covariance matrix elements $q_3$, $q_4$, and $q_5$ are initialized at their steady state values (see \eqref{fpoints_xp} and \eqref{fpoints_special}), the ``stochastic energy'' $E_s = \mathcal{H}$ is conserved in OP evolution, and we will be able to find analytic solutions to the OP equations of motion. \par
Using the above readout values in \eqref{stoch_h_XP} (or, equivalently integrating them out) and fixing the covariance matrix elements to their steady state, gives another form of \eqref{stoch_h_XP}
\begin{equation}
\begin{split}
    &\mathcal{H}^\star(\pmb{p},\pmb{q})=\tfrac{1}{2}\, \pmb{p}^\top \cdot \boldsymbol{\mathsf{B}} \cdot \pmb{p} + \pmb{p}^\top\cdot\boldsymbol{\mathsf{\Omega}}\cdot\pmb{q} - \mathsf{g}.
\end{split}
\label{OP_h_XP}
\end{equation}
The equations of motion for the optimal paths can be found using Hamilton's equations
    $\pmb{\dot{q}}=\partial_{\pmb{p}}\mathcal{H}^\star$, and
    $\pmb{\dot{p}}=-\partial_{\pmb{q}}\mathcal{H}^\star$.
From \eqref{OP_h_XP}, these can be written as 
\begin{equation}
    \begin{split}
        \dot{\pmb{q}}&=\boldsymbol{\mathsf{B}}\cdot\pmb{p}+\boldsymbol{\mathsf{\Omega}}\cdot\pmb{q},\\ \dot{\pmb{p}}&=\boldsymbol{\mathsf{\Omega}}\cdot \pmb{p}.
    \end{split}
    \label{EOM_op_XP}
\end{equation}
The equations for $p_1$ and $p_2$ give rise to sinusoidal solutions. Therefore, the quadrature evolution can be viewed as a forced harmonic oscillator on resonance. When $\boldsymbol{\mathsf{B}}$ is constant (i.e.~the covariance matrix elements are assumed to be time independent with values from \eqref{fpoints_xp} or \eqref{fpoints_special}), we expect to have oscillatory solutions for the quadratures as well, with the coefficients of sine and cosine exhibiting linear time dependence. If we relax the assumption about the covariance matrix elements, however, $q_3$, $q_4$, and $q_5$ will act as some known but time--dependent driving term in the stochastic Hamiltonian. In that case, the stochastic energy is no longer conserved, and the OP equations of motion may be solved numerically.
\subsection{\label{solns_OP}Solution for Optimal Path} Now we discuss the analytical solution of \eqref{EOM_op_XP} when $\tau_1=\tau_2=\mathcal{T}$ (i.e.~in the case of equal measurement strength) for simplicity, with the solution of the general case given in the appendix. The covariance matrix elements take the values \eqref{fpoints_special} and the matrix $\boldsymbol{\mathsf{B}}$ can be evaluated to be $\frac{1}{4\ct}\mathds{1}_2$ i.e. proportional to the 2$\times$2 identity matrix.\par The analytical solution of \eqref{EOM_op_XP} traversing from initial conditions $q_{1i}$ and $q_{2i}$, to the final conditions $q_{1f}$ and $q_{2f}$, over the time interval $\T_f$, is
\begin{equation}
    \begin{split}
        q_1(\T)&=\frac{\T}{\T_f}\{q_{1f}\cos{(\T_f-\T)}-q_{2f}\sin{(\T_f-\T)}\}\\&+\Big(1-\frac{\T}{\T_f}\Big)(q_{1i}\cos{\T}+q_{2i}\sin{ \T}),\\
         q_2(\T)&=\frac{\T}{\T_f}\{q_{1f}\sin{(\T_f-\T)}+q_{2f}\cos{(\T_f-\T)}\}\\&+\Big(1-\frac{\T}{\T_f}\Big)(-q_{1i}\sin{\T}+q_{2i}\cos{\T}),\\
         p_1(\T)&=\frac{4\ct}{\T}\{q_{1f}\cos{(\T_f-\T)}-q_{2f}\sin{(\T_f-\T)}\\&-(q_{1i}\cos{\T}+q_{2i}\sin{\T})\},\\
         p_2(\T)&=\frac{4\ct}{\T_f}\{q_{1f}\sin{(\T_f-\T)}+q_{2f}\cos{(\T_f-\T)}\\&-(-q_{1i}\sin{\T}+q_{2i}\cos{\T})\}.
    \end{split}
    \label{OP_sol_equal}
\end{equation}
As discussed previously, the solutions for $p_1$ and $p_2$ are sinusoidal while $q_1$ and $q_2$ are sinusoidal with coefficients growing linearly with $\T$. Note, there is no dependence on strength $\ct$ of measurement in $q_1(\T)$ and $q_2(\T)$. The strength only appears in the probability of the most-likely path. We see that the boundary conditions of the position and momentum define a unique solution. As a result, ``multipath solutions'' \cite{LewalleMultipath, NaghilooCaustic, LewalleChaos}\footnote{Some system dynamics allow for multiple OP solutions connecting particular boundary conditions, in close analogy with the formation of caustics in ray solutions of optical problems. The uniqueness of the solutions above is a direct proof that this phenomenon \emph{cannot} occur in the present case.} do not exist in this system, and we may further understand that every OP solution is in fact a most-likely path, simplifying the interpretation of the solutions considerably.\par The action is often treated as a generating function which transforms between different dynamical variables at different times \cite{Landau1976Mechanics, BookArnoldClassical, goldstein:mechanics}. The relation 
$\pmb{p}_f=\partial_{\pmb{q}_f} \mathcal{S}$ is of particular interest in our present case, as it indicates that the globally most likely OP is the one that terminates at $\pmb{p}_f=0$ \cite{LewalleMultipath}. Its final coordinates
\begin{equation}
\begin{split}
    \mathcal{Q}_{1f}&=q_{1i}\cos{\T_f}+q_{2i}\sin{\T_f},\\
     \mathcal{Q}_{2f}&=-q_{1i}\sin{\T_f}+q_{2i}\cos{\T_f},
    \end{split}
    \label{Qfs}
\end{equation}
denote the globally most--likely final state under the measurement dynamics, given the initial state $\pmb{q}_i$ at $\T = 0$.
This particular MLP is a circular path in the phase space following the unitary dynamics alone, and in fact has $\pmb{p} = 0$ over its entire evolution in our present example.  In our system, the globally most-likely path is also the average path over all possible stochastic trajectories. Other choices of $p_1$ and $p_2$ correspond to post--selection on other possible (but less--likely) final states allowed under the diffusive measurement dynamics.
\begin{figure}
	\includegraphics[width=0.45\textwidth,trim = {0 0 0 60}, clip]{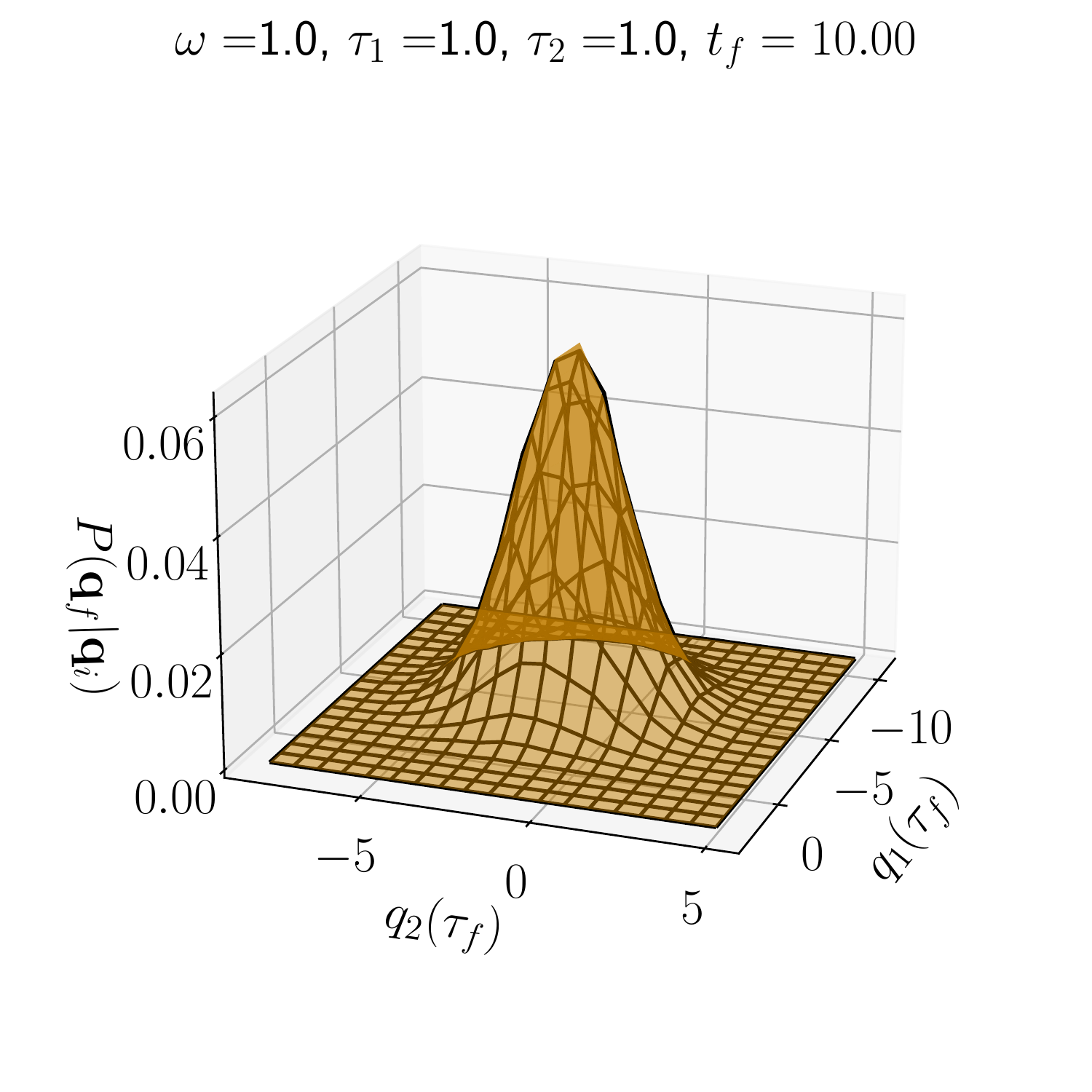}\\ \vspace*{-.5cm}
	\includegraphics[width=0.45\textwidth,trim = {0 0 0 30}, clip]{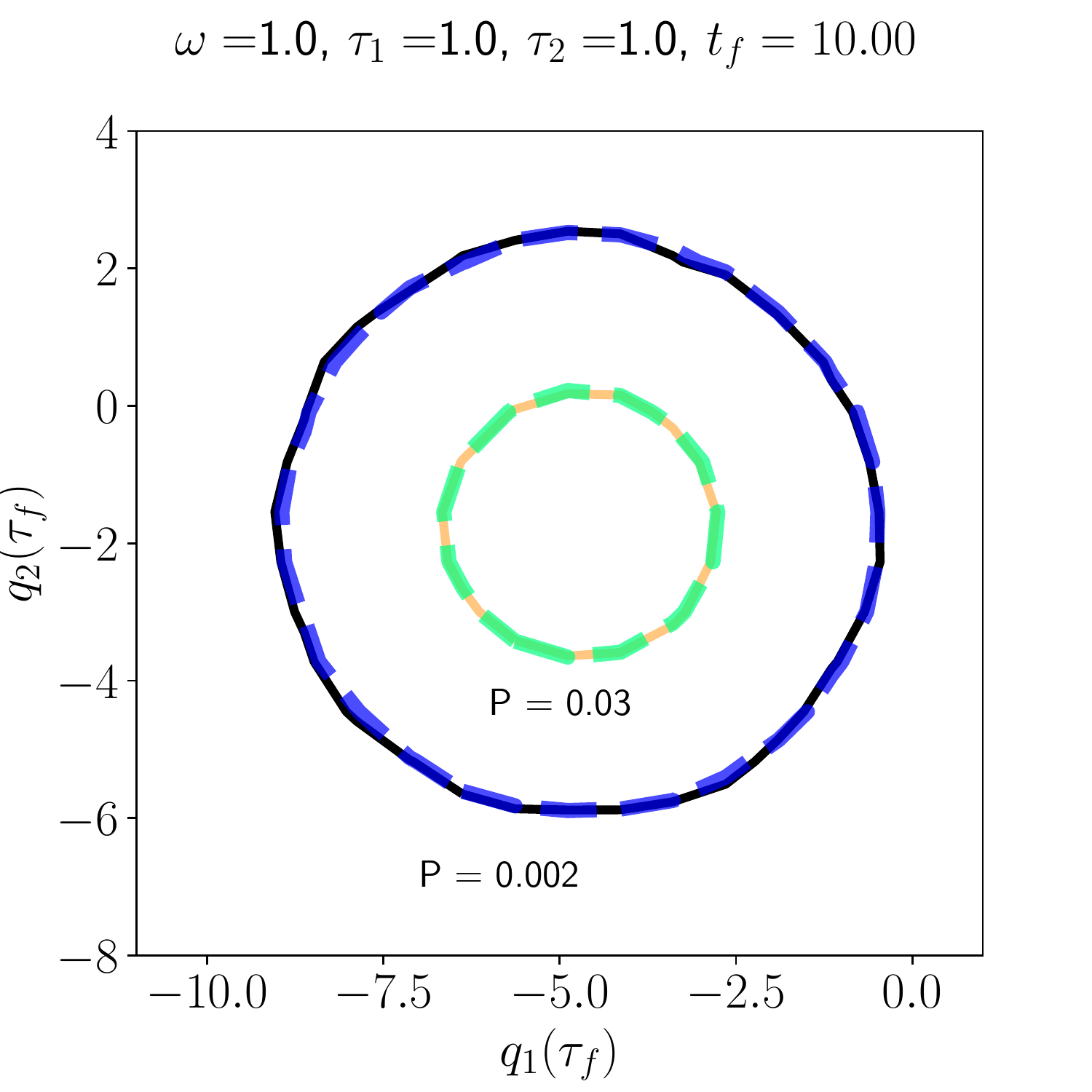}
	\begin{tikzpicture}[overlay]
	\node[] at (-8.,14.5) {(a)};
	\node[] at (-8,7.5) {(b)};
	\end{tikzpicture}
	\caption{Probability density of final state at $\T_f=10.0$ with initial conditions $q_1(0)=3$, $q_2(0)=4$ for $\tau_1=\tau_2=\ct=1.0$. Panel (a) shows the 3-dimensional plot of the probability density function as a function of final quadratures. The black wireframe corresponds to the results from simulations of 100,000 stochastic trajectories. The solid orange surface corresponds to analytical probability calculated from \eqref{P_equal}. Panel (b) shows the contours with the values $P(\pmb{q}_f|\pmb{q}_i)=0.03$ (inner circle), and 0.002 (outer circle). Thick dashed contours correspond to the analytical expression \eqref{P_equal} and the thinner solid  line corresponds to probability calculated from the simulations. The exact matching with the simulations in both figures confirm the validity of the path integral calculation shown in \ref{sec_prob_density}  and Appendix \ref{appA4}.\vspace{-.4cm}} 
	\label{prob_density_3}
\end{figure}
\section{\label{sec_prob_density}Final state probability density} 
In this section we exactly calculate  $P(\pmb{q}_f|\pmb{q}_i)$, i.e.~the probability density of the final state $\pmb{q}_f(\T_f)$,  starting from the initial state $\pmb{q}_i$. We can calculate this quantity by integrating the probability density  \eqref{action_integral} over all possible quadratures and readouts,
\begin{equation}
    \begin{split}
    P&(\pmb{q}_f|\pmb{q}_i)=\int\mathcal{D}\pmb{p}\mathcal{D}\pmb{q}\mathcal{D}\pmb{r}\:e^\mathcal{S}\\&=\int\mathcal{D} \pmb{p}\mathcal{D}\pmb{q}\mathcal{D}\pmb{r}\: \exp\left[\int_{0}^{\T_f}d\T(-\pmb{p}\cdot\pmb{\dot{q}}+\mathcal{H}(\pmb{p},\pmb{q},\pmb{r}))\right].
    \end{split}
    \label{pqfqi}
\end{equation}
Using \eqref{stoch_h_XP} and integrating w.r.t.~readouts (see Appendix \ref{appA4} for details) we get 
\begin{equation}
\begin{split}
    P(\pmb{q}_f|\pmb{q}_i)=\mathcal{N}_1\int\mathcal{D} \pmb{p}\mathcal{D}\pmb{q}\: e^{-\int_{0}^{\T_f}d\T\left(\pmb{p}\cdot\dot{\pmb{q}}-\frac{1}{2}\mathbf{p}^\top\cdot\boldsymbol{\mathsf{B}}\cdot\mathbf{p}-\mathbf{p}^\top\cdot\boldsymbol{\mathsf{\Omega}}\cdot \mathbf{q}+\mathsf{g}\right)},
\end{split}
    \label{p_rest_int_1}
\end{equation} with $\mathcal{N}_1$ as a constant only dependent on $\T_f$ and the measurement strengths. The functional integration over $\pmb{p}$ and $\pmb{q}$ can be carried out to find
\begin{equation}
    P(\pmb{q}_f|\pmb{q}_i)=\mathcal{N}\exp{-\frac{1}{2}\int_0^{\T_f}d\T\: \mathbf{p}_{\mathsf{OP}}^\top\cdot\boldsymbol{\mathsf{B}}\cdot\mathbf{p}_{\mathsf{OP}}},
    \label{P_expr1_1}
\end{equation}
where $\mathcal{N}$ is the normalization constant, and $\mathbf{p}_{\mathsf{OP}}$ denotes the analytical solution for the optimal path momenta with given boundary conditions. For the equal measurement strengths, this reduces to 
\begin{equation}
    P(\pmb{q}_f|\pmb{q}_i)=\frac{2\ct}{\pi \T_f}\exp{-\frac{2\ct}{\T_f}(\Delta Q_1^2+\Delta Q_2^2)},
    \label{P_equal}
\end{equation}
where $\Delta Q_1=q_{1f}-\mathcal{Q}_{1f}$ and $\Delta Q_2=q_{2f}-\mathcal{Q}_{2f}$. We see that the solution depends on the measurement time $\ct$ relative to the duration $\T_f$ and is Gaussian in $\Delta Q_1$ and $\Delta Q_2$. The variance is $\T_f/\ct$. This is consistent with the fact that stronger measurement (smaller values of $\ct$) leads to larger diffusion in the trajectories due to measurement backaction. Fig.~\ref{prob_density_3} compares these analytical results from path-integral calculations with 100,000 trajectories generated from \eqref{q_stoch}, confirming the above calculation.
\section{\label{energetics}Optimal Path Energy}
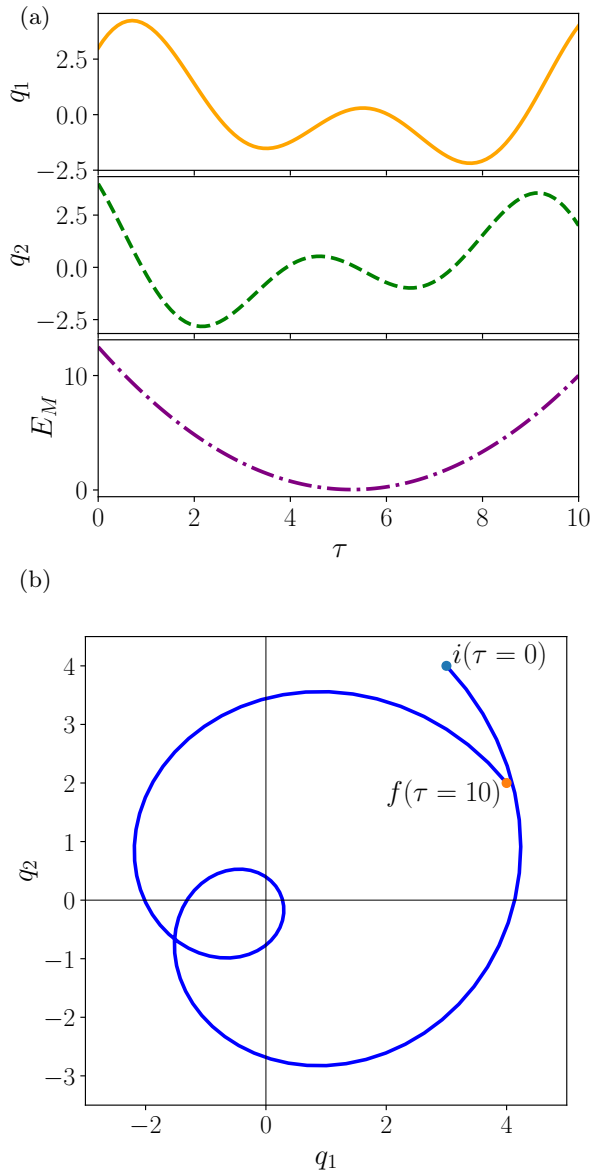
\begin{figure}
	\includegraphics[width=.985\linewidth, trim = {0 0 0 30}, clip]{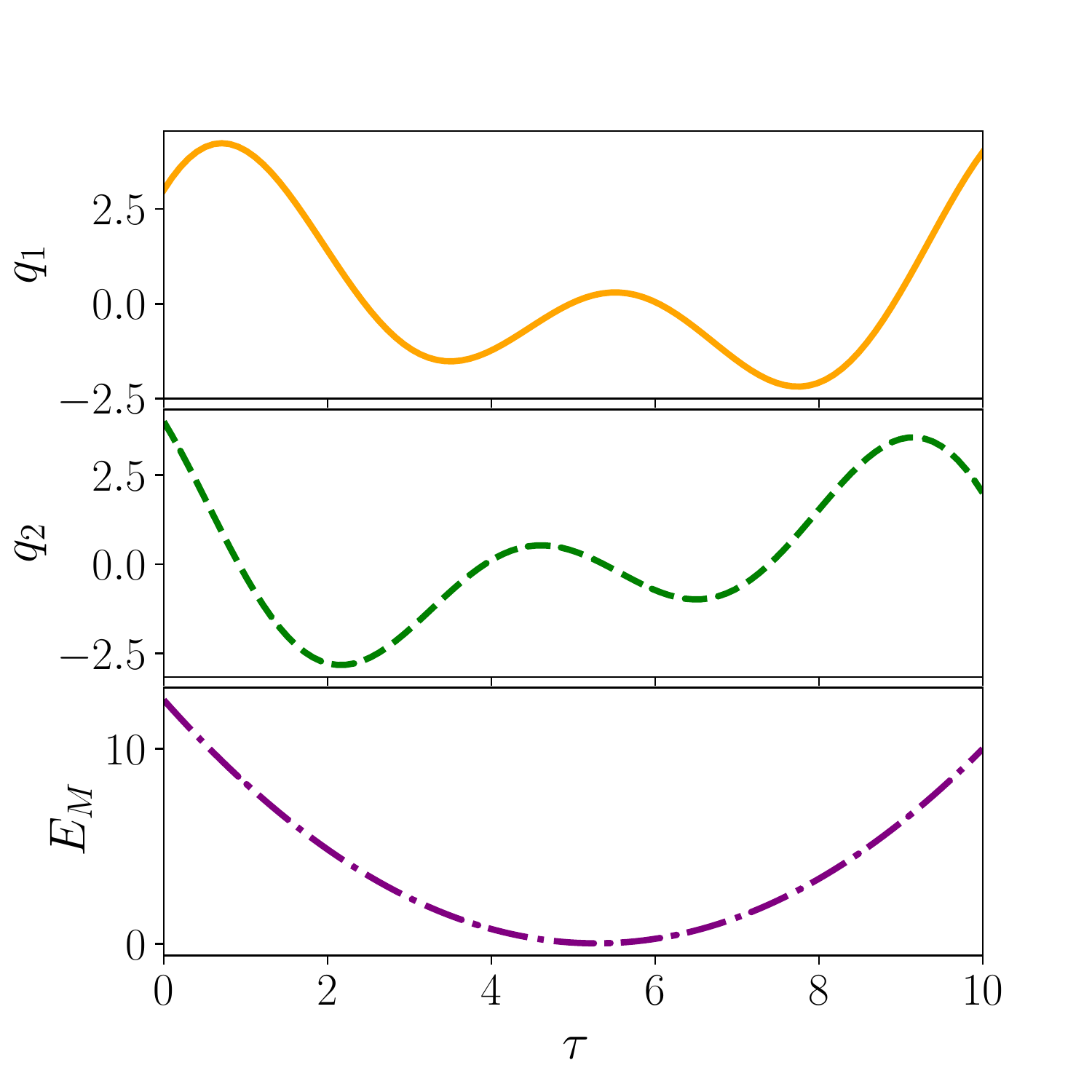} \\
	\includegraphics[width=.955\linewidth, trim = {0 0 0 12}, clip]{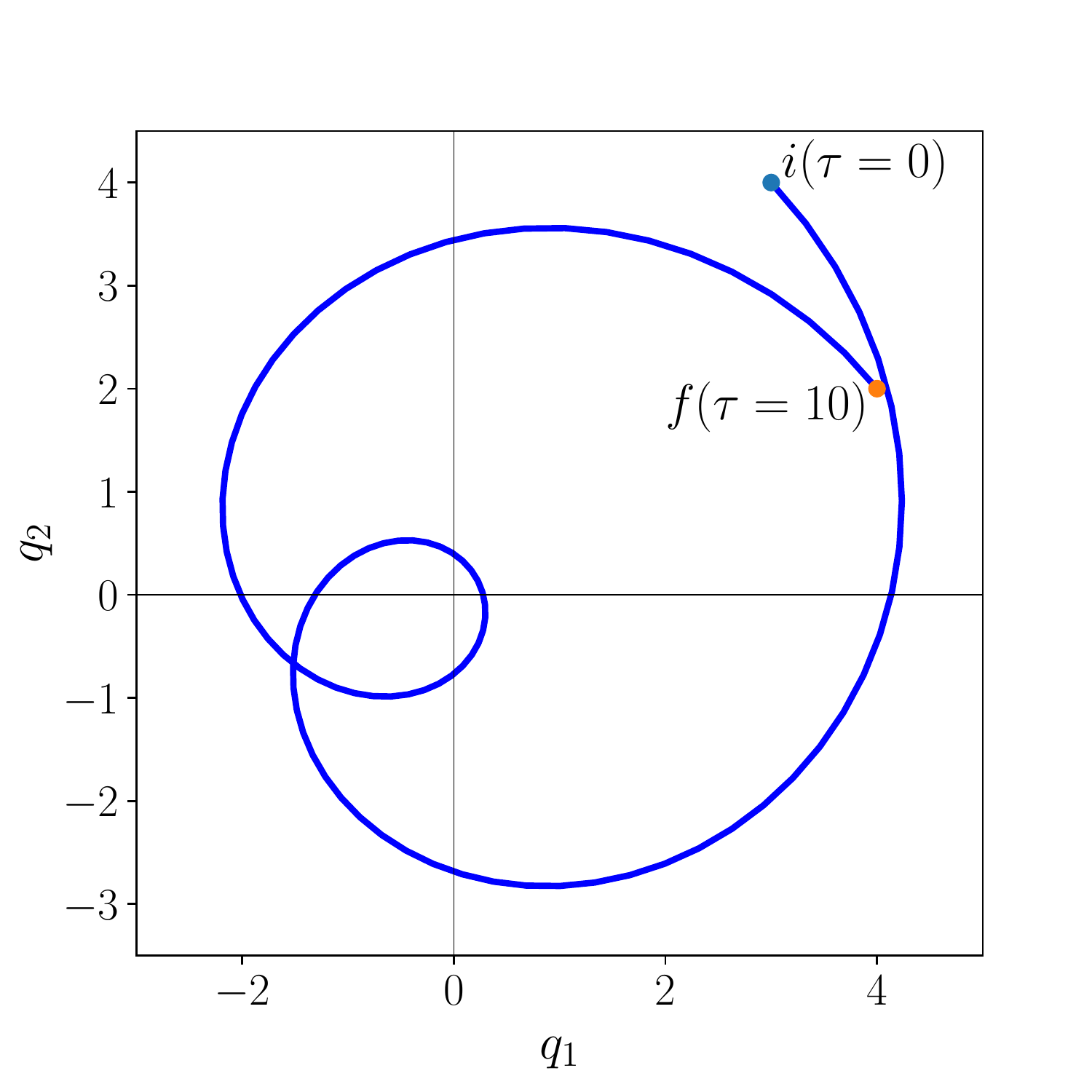} 
	\begin{tikzpicture}[overlay]
	\node[] at (-8,15.5) {(a)};
	\node[] at (-8,8) {(b)};
	\end{tikzpicture}
	\caption{(a) and (b) show the time evolution of the quadratures for boundary conditions $q_1(0)=3$, $q_2(0)=4$, $q_1(\T_f=10)=4$, and $q_2(\T_f=10)=2$. In panel (a) $q_1$ (top), $q_2$ (middle), and the mechanical energy $E_M$ (botttom panel) from \eqref{e_mech_equal}  are shown for a sample optimal path for the equal measurement strength case ($\tau_1 = \tau_2=\ct=1$). The sinusoidal evolution of the quadratures is due to the unitary dynamics while the change in amplitude is due to the diffusive dynamics. In panel (b) we plot the sample optimal path in the quadrature space.}
	\label{trajectory_2}
\end{figure}
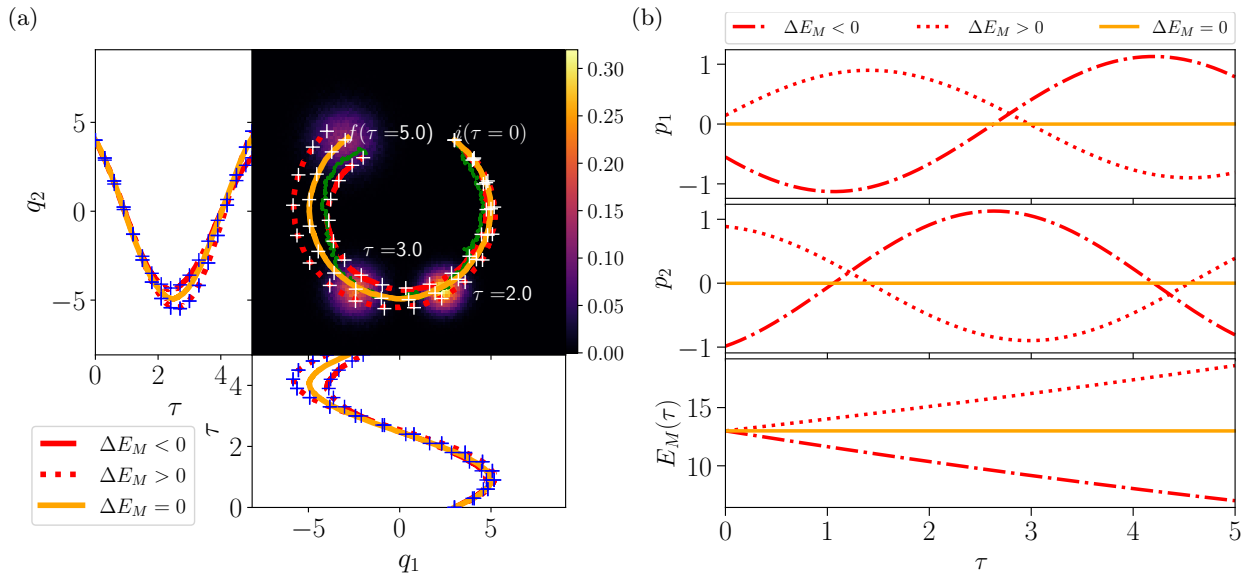
\begin{figure*}
    \includegraphics[width=0.45\textwidth,trim = {0 0 0 50}, clip]{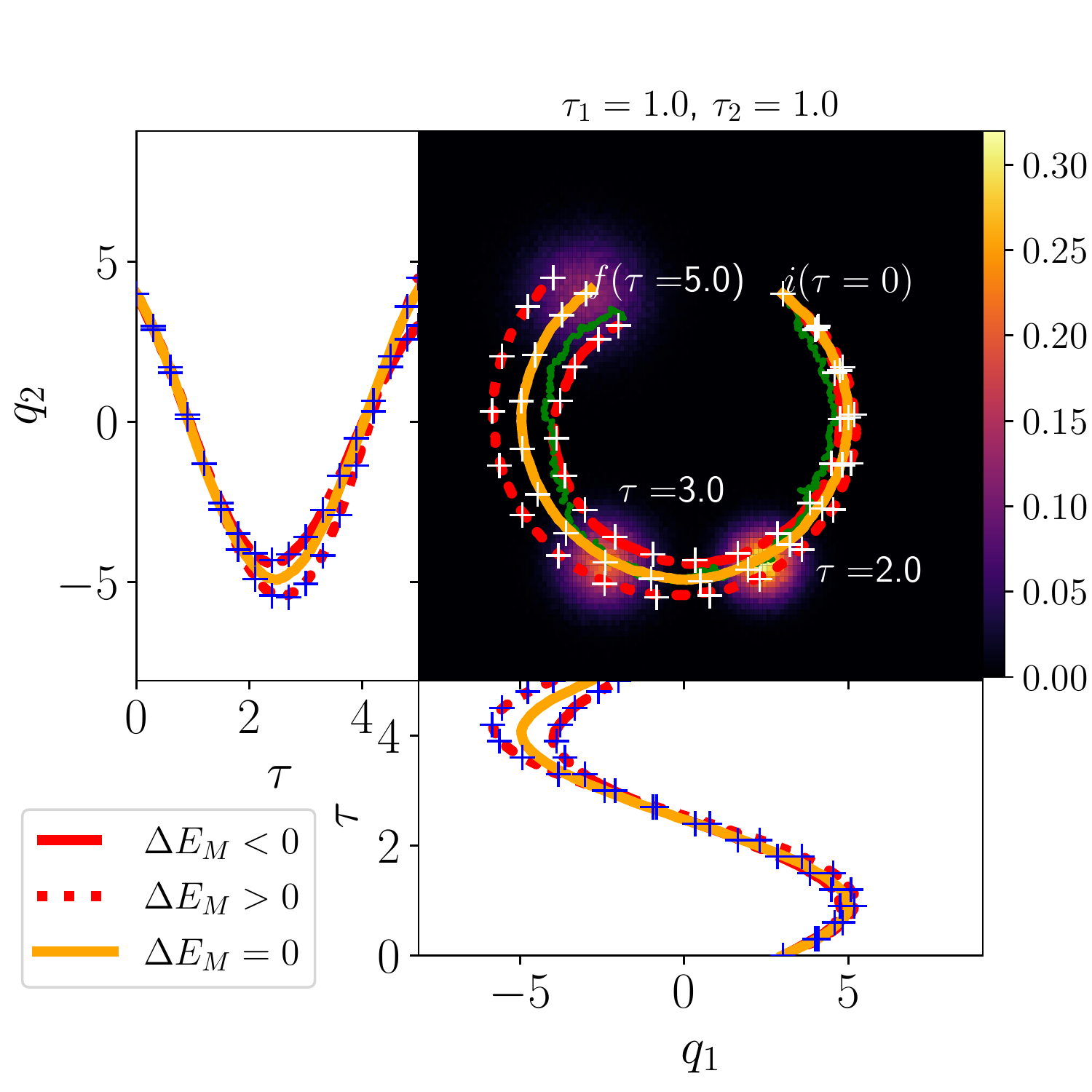}
	\includegraphics[width=0.45\textwidth,trim = {0 0 0 25}, clip]{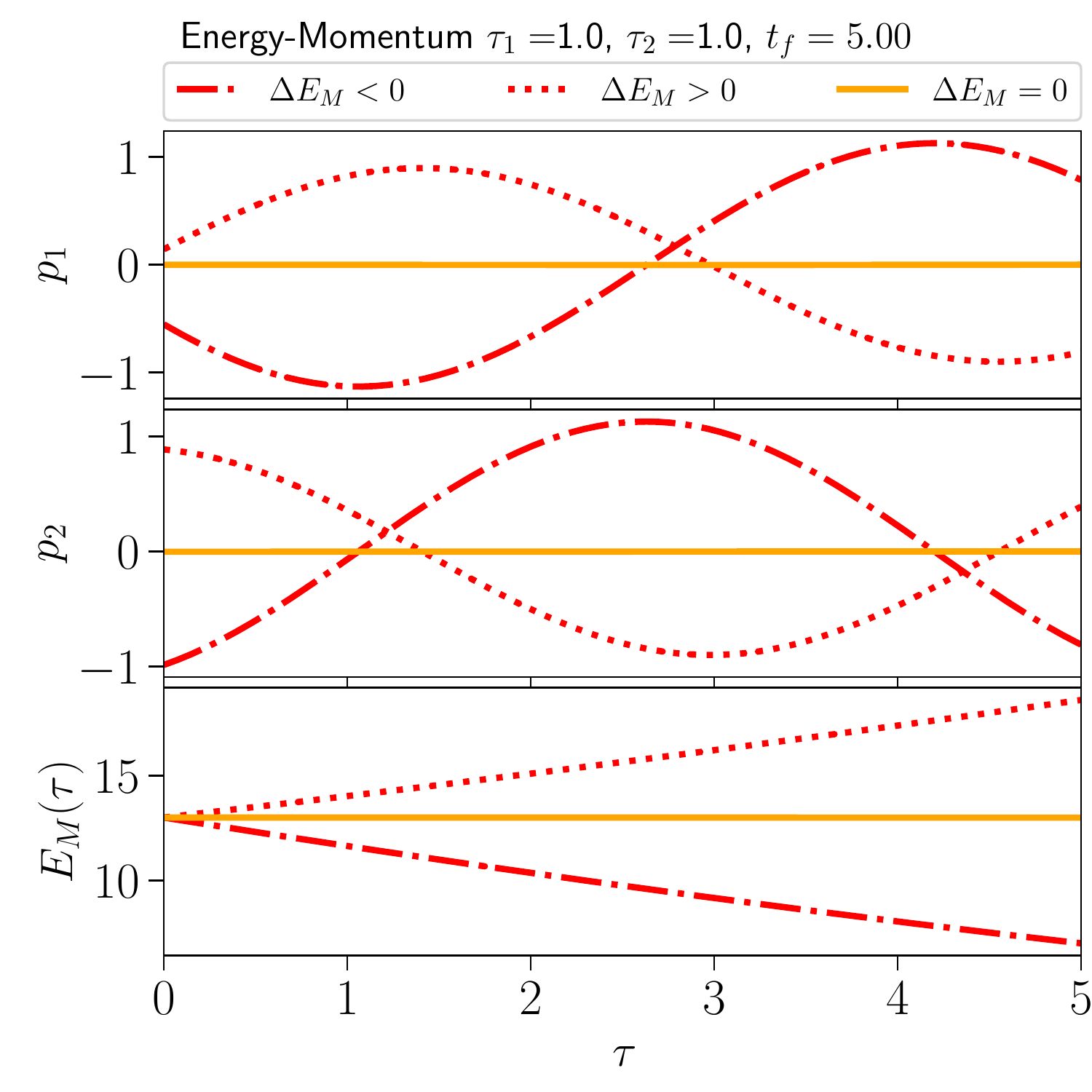}
	\begin{tikzpicture}[overlay]
	\node[] at (-16.3,7.5) {(a)};
	\node[] at (-8,7.5) {(b)};
	\end{tikzpicture}
	\caption{(a) We illustrate the evolution of trajectories and optimal paths starting from $q_1(0)=3$, $q_2(0)=4$ for $\tau_1=\tau_2=\ct=1.0$ till $\T_f=5$. Note $\T=2\pi$ corresponds to a full energy conserving cycle. Yellow solid line corresponds to the average of 50 clustered trajectories for end points on $p_1=p_2=0$ circle ($q_1(5.0)=-2.98$, $q_2(5.0)=4.01$). The red dotted line corresponds to the average of 50 trajectories which lead to an increase in the mechanical energy of the system ($q_1(5.0)=-4.00$, $q_2(5.0)=4.50$). Red dot-dashed line is for average of trajectories with decrease in the mechanical energy of the system ($q_1(5.0)=-2.00$, $q_2(5.0)=3.00$). In the top right panel we show the optimal paths in quadrature space. White `+' shows the analytical solutions corresponding to the three boundary conditions mentioned and the  green line is a sample stochastic trajectory. Three histograms of the quadratures for three different times ($\T$=2.0, $\T$=3.0, $\T$=5.0 superimposed) are also shown along with a colorbar on the right depicting the probability density values (see Section \ref{sec_prob_density}) calculated from the simulations. The histograms show diffusion of the trajectories starting from the same initial point. Individual trajectories are not smooth but their post-selected clustered averages give us the smooth analytical optimal paths exactly. In top left and bottom panels we compare the time dependence of individual quadratures between the averaged stochastic trajectories and the analytical solution of the three optimal paths (blue `+'). The analytical time dependence show excellent agreement with averaged results, supporting our interpretation of the optimal path. Panel (b) shows the time evolution of the conjugate momenta and the mechanical energy for the three optimal paths shown in panel (a). Top two panels show $p_1$ and $p_2$ to be both sinusoidal for $\Delta E_M\neq 0$ and 0 for the energy conserving path. The bottom most panel shows $E_M$ to be increasing/decreasing/constant depending on the optimal path.}
	\label{PS_traj_29}
\end{figure*}
The solution \eqref{OP_sol_equal} can also  be used to understand the energetics of measuring the oscillator, opening up the possibility of its adaptation in making quantum measurement engines/refrigerators \cite{PhysRevLett.120.260601, jordan_quantum_2020} and for feedback cooling \cite{rossi_measurement-based_2018,wilson_measurement-based_2015,krause2015optical,PhysRevLett.99.017201,kleckner_sub-kelvin_2006,PhysRevLett.99.160801}.
Using \eqref{OP_h_XP} and \eqref{OP_sol_equal}, the expectation value of mechanical energy of the oscillator on a general optimal path is $E_M(\T)=$
\begin{equation}
    \frac{1}{2}+\frac{1}{2}\Big\{\Big(\mathcal{Q}_{1f}+\frac{\T}{\T_f}(q_{1f}-\mathcal{Q}_{1f})\Big)^2+\Big(\mathcal{Q}_{2f}+\frac{\T}{\T_f}(q_{2f}-\mathcal{Q}_{2f})\Big)^2\Big\}.
    \label{e_mech_equal}
\end{equation}
The mechanical energy evolves quadratically in time. The change in energy of the system is provided by the measurement. For $q_{1f}=\mathcal{Q}_{1f}$ and $q_{2f}=\mathcal{Q}_{2f}$ (i.e.~final points on globally most-likely optimal path) the mechanical energy stays constant with time. \par
The expression \eqref{e_mech_equal} also restricts the values of final mechanical energies  attainable during the measurement process with a certain measurement final time $\T_f$ and boundary conditions. To illustrate this effect, Fig.~\ref{trajectory_2}(a) shows the quadratures and mechanical energy as functions of time for a sample optimal path. The quadratures are sinusoidal with time dependent amplitude. The mechanical energy is parabolic, i.e.~measurement first takes away from, but then provides energy to, the system. However, for the conditions chosen, the mechanical energy can only take the values lying on the purple dot-dashed curve. The analytical expression of the mechanical energy \eqref{e_mech_equal} therefore tells us which states to post-select if we want to add to or extract energy from the system.\par  
Fig.~\ref{trajectory_2}(b) shows the same optimal path  in quadrature space. The change in amplitude due to diffusive dynamics leads to spiral OPs instead of circular paths we see in unitary dynamics.  The distance from the origin determines the mechanical energy of the oscillator and evidently, for the sample optimal path, the mechanical energy first decreases then increases with an overall decrease in final time. The stochastic energy \cite{chantasri2013action,PhysRevA.92.032125,Areeya_Thesis,LewalleMultipath,LewalleChaos} is 
\begin{equation}
\begin{split}
    E_s&=\frac{2\ct}{\T_f^2}((q_{1f}-\mathcal{Q}_{1f})^2+(q_{2f}-\mathcal{Q}_{2f})^2)\\&+\frac{4\ct}{\T_f}(q_{1f}\mathcal{Q}_{2f}-q_{2f}\mathcal{Q}_{1f})-\frac{1}{2\ct},
\end{split}
    \label{Es_equal}
\end{equation}
which is a constant of motion. 

\par Although \eqref{e_mech_equal} tells us that the mechanical energy does not depend on the value of $\ct$, the probability of achieving a final state (i.e.~post-selection) does. This implies that arbitrarily large changes in energy are possible, but that larger changes correspond to rarer events.  Therefore both \eqref{e_mech_equal} and \eqref{P_equal} will come into play while designing quantum measurement engines or refrigerators  based on this system \cite{PhysRevLett.120.260601, jordan_quantum_2020}, as these engines rely on taking energy from the measurement process.
\section{\label{OP_traj}Connection between optimal path and stochastic trajectories}
Optimal paths for the Gaussian states can be defined as the most likely trajectory connecting a pair of boundary conditions. A complementary way to look at this is to post-select stochastic trajectories with particular boundary conditions and average the subset of them that are closest to each other since we expect the trajectories to follow the MLP closely.\par Following the method laid out in  \cite{NaghilooCaustic, weber_mapping_2014}, we post--select 500 simulated trajectories according to the given boundary conditions. Then identify the subset of fifty trajectories that have the least distance between each other over their evolution. Such a grouping identifies a cluster of paths, whose centroid is approximated by averaging the clustered trajectories.   Fig.~\ref{PS_traj_29} shows a comparison between these MLPs extracted from simulated data, and the analytic MLP \eqref{OP_sol_equal}, for three different boundary conditions.  The time dependence of $q_1$ and $q_2$ and the trajectory in quadrature space match exactly. Fig.~\ref{PS_traj_29}(a) also shows histograms of the quadratures for 100,000 trajectories at times $\T=2$, 3, and 5.  The spread of the histograms limits the extent to which energy can be added or withdrawn from the system via diffusive measurements with significant probability in a certain amount of time.\par   
 Fig.~\ref{PS_traj_29}(b) shows the analytical solution of $p_1$, $p_2$ (momenta conjugate to $q_1$ and $q_2$) and $E_M$ as functions of time for the three different boundary conditions previously shown in Fig.~\ref{PS_traj_29}(a). Mechanical energy increases or decreases due to measurement for the choice of boundary conditions. As the sample boundary conditions $\pmb{q}_f$ are very close to $\pmb{\mathcal{Q}}_f$, the energy changes are almost linear in time.

\section{\label{disc}Discussion}
We have generalized the CDJ path integral formalism \cite{Areeya_Thesis,PhysRevA.92.032125,chantasri2013action} to harmonic oscillators in general Gaussian states undergoing continuous and weak simultaneous position and momentum measurements. The characteristic function gives us the evolution equations for the expectation values of position and momentum, and the covariance matrix elements. While the first two evolve stochastically, the covariance matrix elements follow a set of three  deterministic self-contained ordinary differential equations, describing purification of the state by measurement. Using the CDJ stochastic path integral formalism, we find probabilities for diffusive trajectories of the system in terms of a stochastic action. The most likely paths, found from extremizing the action, take the form of a resonantly driven oscillator. This leads to the change in amplitude and therefore mechanical energy of the oscillation due to measurement. The optimal path description and the time dependent expression  of the mechanical energy  have promising applicability for designing feedback control of quantum oscillators. For example, extracting energy through measurement for charging a quantum battery \cite{2020arXiv201200350M} or  cooling of mechanical resonator \cite{rossi_measurement-based_2018} provide exciting avenues for the incorporation of our results. For this particular system, we also successfully establish the interpretation of optimal paths as the average of clustered post-selected  stochastic trajectories.\par In this work we were able to evaluate the final state probability density under continuous measurements by making exact functional integrals of the stochastic path integral. For the equal measurement strength case, this probability density is  Gaussian centered around the globally most likely path. Generalization of all results to unequal measurement strength, including position only measurement, are given in the appendices. Although restricted to Gaussian states, these results can be useful for a vast range of  experiments due to the ubiquity  of such states. Our analysis offers avenues for future developments on quantum trajectories  of continuous variable systems.
\section*{\label{sec:level1}Acknowledgement}
We thank Sreenath Manikandan for providing insight throughout the project. This work has been supported by NSF grant no.~DMR-1809343, US Army Research Office grant no.~W911NF-18- 10178, and the John Templeton Foundation grant no.~61835.  
\appendix
\section{\label{appA0}Gaussian States}
The characteristic function of the density matrix $\hat{\rho}$ is defined as \cite{wang2007quantum} 
\begin{equation}
    \chi(\pmb{\xi}) = \text{Tr}(\rho e^{i\pmb{\xi}^\top\cdot \pmb{\hat{\mathcal{R}}}}).
    \label{char_fn}
\end{equation}
Here $\pmb{\hat{\mathcal{R}}}$ is a vector of operators given by $(\hat{X}, \hat{P})^\top$ and $\pmb{\xi}$ is a vector in $\mathbb{R}^2$. For a Gaussian state, the characteristic function takes the form \cite{wang2007quantum}:
\begin{equation}
    \chi(\pmb{\xi}) = \exp\left(-\frac{1}{4}\pmb{\xi}^\top\cdot\Gamma\cdot\pmb{\xi}+i\pmb{\mu}^\top\cdot\pmb{\xi}\right).
    \label{char_gaussian}
\end{equation}
$\Gamma$ is a $2\times 2$ positive-definite real symmetric matrix\cite{PhysRevA.49.1567}, also known as the covariance matrix
\begin{equation}
    \Gamma=\begin{pmatrix}
q_3 & q_4\\
q_4 & q_5
\end{pmatrix}    \label{cov_matrix}
\end{equation}
and real column vector $\pmb{\mu}=(q_1,\:\:q_2)^\top$.  A Gaussian state is characterized completely by the matrices $\Gamma$ and $\mu$. For pure states $\det \Gamma=1$.
\section{\label{appA1}Covariance Matrix Evolution}The three equations of \eqref{q_deter} give the deterministic evolution of the covariance matrix elements. For all values of $\tau_1$, and $\tau_2$ they have one and only fixed point \eqref{fpoints_xp}. We will now prove that the solutions to these three equations for any initial values of covariance matrix elements tend to the fixed points for $\T\to\infty$.\\ Consider the determinant of the covariance matrix \eqref{cov_matrix}, $A=q_3q_5-q_4^2$. Now, using the Heisenberg uncertainty relation \cite{sakurai_napolitano_2017}, $\expval{(\Delta \hat{X})^2}\expval{(\Delta \hat{P})^2}$
\begin{equation}
    \geq\frac{1}{4}|\expval{[\Delta\hat{X},\Delta\hat{P}]}|^2+\frac{1}{4}|\expval{\{\Delta\hat{X},\Delta\hat{P}\}}|^2
    \label{uncertainty}
\end{equation}
leads to 
\begin{equation}
    A=q_3q_5-q_4^2\geq1.
    \label{a_ineq}
\end{equation}
From \eqref{q_deter}, we get 
\begin{equation}
    \frac{dA}{d\T}=-\Big(\frac{q_3}{2\tau_1}+\frac{q_5}{2\tau_2}\Big)(A-1).
    \label{det_evol}
\end{equation}
As $q_3$ and $q_5$ are non-negative, \eqref{a_ineq} and \eqref{det_evol} show that the determinant is always decreasing with time. Integrating \eqref{det_evol} gives
\begin{equation}
    \ln{\abs{\frac{A-1}{A_i-1}}}=-\int_0^\T\Big(\frac{q_3}{2\tau_1}+\frac{q_5}{2\tau_2}\Big)d\T^\prime.
    \label{integral}
\end{equation}
Using an inequality between the arithmetic and the geometric mean, 
\begin{equation}
    \frac{q_3}{2\tau_1}+\frac{q_5}{2\tau_2}\geq\sqrt{\frac{q_3q_5}{\tau_1\tau_2}}\geq\sqrt{\frac{1}{\tau_1\tau_2}},
    \label{ineq}
\end{equation}
we find that
\begin{equation}
    \abs{\frac{A-1}{A_i-1}}\leq \exp{-\frac{\T}{\sqrt{\tau_1\tau_2}}}.
    \label{a_ineq_t}
\end{equation}
Therefore $A\to 1$ as $\T\to\infty$ for any valid $A_i$ implying that the system goes towards a pure states for time continuous weak simultaneous position-momentum measurements. We assume $A_i$ to be the determinant value ($\geq 1$) at $\T=0$. If $A_i =1$, the determinant stays at 1 for all $\T$.\par Now consider the variables $x$, $y$ and $z$ defined as:
\begin{equation}
        q_3=\tilde{q}_3+x,\:\:\:q_4=\tilde{q}_4+y,\:\:\:\:q_5=\tilde{q}_5+z.
    \label{xyz}
\end{equation}
From \eqref{q_deter} the evolution of these variables can be determined:
\begin{equation}
    \begin{split}
        \frac{dx}{d\T}&=-\frac{\tilde{q}_3}{\tau_1}x+\Big(2-\frac{\tilde{q}_4}{\tau_2}\Big)y-\Big(\frac{x^2}{2\tau_1}+\frac{y^2}{2\tau_2}\Big),\\
        \frac{dy}{d\T}&=-\Big(1+\frac{\tilde{q}_4}{2\tau_1}\Big)x-\Big(\frac{\tilde{q}_3}{2\tau_1}+\frac{\tilde{q}_5}{2\tau_2}\Big)y\\&+\Big(1-\frac{\tilde{q}_4}{2\tau_2}\Big)z-\Big(\frac{xy}{2\tau_1}+\frac{yz}{2\tau_2}\Big),\\
        \frac{dz}{d\T}&=-\Big(2+\frac{\tilde{q}_4}{\tau_1}\Big)y-\frac{\tilde{q}_5}{\tau_2}z-\Big(\frac{y^2}{2\tau_1}+\frac{z^2}{2\tau_2}\Big).
    \end{split}
    \label{xyz_evol}
\end{equation}
As $A\to 1$ for $\T \to \infty$, 
\begin{equation}
    \lim_{t\to\infty}(\tilde{q}_5 x-2\tilde{q}_4 y+\tilde{q}_3 z+xz-y^2)=0.
    \label{limit}
\end{equation}
\eqref{xyz_evol} gives
\begin{equation}
    \frac{d}{d\T}(xz-y^2)=-\frac{1}{2}\Big(\frac{2\tilde{q}_3+x}{\tau_1}+\frac{2\tilde{q}_5+z}{\tau_2}\Big)(xz-y^2).
    \label{xz_y2_evol}
\end{equation}
Integrating yields
\begin{equation}
    \abs{\frac{xz-y^2}{x_iz_i-y_i^2}}=\exp{-\int_{0}^\T\frac{1}{2}\Big(\frac{\tilde{q}_3+q_{3}}{\tau_1}+\frac{\tilde{q}_5+q_{5}}{\tau_2}\Big)d\T^\prime}.
    \label{xz_y2_sol}
\end{equation}
The denominator on the right hand side corresponds to initial values. Again applying the inequality between arithmetic and geometric mean, we get:
\begin{equation}
    \abs{\frac{xz-y^2}{x_iz_i-y_i^2}}\leq\exp{-\sqrt{\frac{2}{\tau_1\tau_2}}\T}.
    \label{xz_y2_lim}
\end{equation}
While \eqref{a_ineq_t} gives \eqref{limit}, \eqref{xz_y2_lim} proves that the nonlinear part of \eqref{limit} i.e. $xz-y^2$ also goes to 0. We denote the $\T\to\infty$ limits of $x$, $y$ and $z$ by $x_f$, $y_f$ and $z_f$ respectively. From \eqref{limit} and the fact that $\tilde{q}_3 \tilde{q}_5-\tilde{q}_4^2=1$, we get
\begin{equation}
    (\tilde{q}_5x_f-\tilde{q}_3z_f)^2+4y_f^2=0.
    \label{xfyfzf}
\end{equation}
\eqref{xfyfzf} and \eqref{limit} gives $x_f, y_f, z_f =0$. Hence, we have provided a proof for the statement that at $\T\to\infty$ the values of $q_3, q_4, q_5$ tends to the fixed points of \eqref{q_deter}.

\section{\label{appA2}Properties of the Fixed Point}We can linearize \eqref{xyz_evol} around the fixed point to to get the following:
\begin{equation}
    \frac{d}{d\T}\begin{pmatrix}x\\y\\z\end{pmatrix}=\mathcal{M}\begin{pmatrix}x\\y\\z\end{pmatrix},
    \label{lin_M}
\end{equation}
with the matrix 
\begin{equation}
\begin{split}
    \mathcal{M}=\begin{pmatrix} -\frac{\tilde{q}_3}{\tau_1} &\Big(2-\frac{\tilde{q}_4}{\tau_2}\Big)&0\\[8pt] -\Big(1+\frac{\tilde{q}_4}{2\tau_1}\Big)& -\Big(\frac{\tilde{q}_3}{2\tau_1}+\frac{\tilde{q}_5}{2\tau_2}\Big)&\Big(1-\frac{\tilde{q}_4}{2\tau_2}\Big) \\[8pt] 0& -\Big(2+\frac{\tilde{q}_4}{\tau_1}\Big)& -\frac{\tilde{q}_5}{\tau_2} \end{pmatrix}.
\end{split}
\label{lin_matrix}
\end{equation}
The eigenvalues of the \eqref{lin_matrix} are,
\begin{equation}
    \begin{split}
        \lambda_0&=-\Big(\frac{\tilde{q}_3}{2\tau_1}+\frac{\tilde{q}_5}{2\tau_2}\Big),\\
        \lambda_{+,-}&=-\Big(\frac{\tilde{q}_3}{2\tau_1}+\frac{\tilde{q}_5}{2\tau_2}\Big),\\&\pm i \sqrt{\frac{1}{4}\Big(\frac{\tilde{q}_3}{2\tau_1}+\frac{\tilde{q}_5}{2\tau_2}\Big)^2+4-\frac{1}{\tau_1\tau_2}}.
    \end{split}
    \label{eigen}
\end{equation}
As the real parts of the eigenvalues are negative, the fixed point is stable (with node in one direction and inward spirals along the other two directions). 
\section{\label{appA3}General Solution for Optimal Path}
Solving \eqref{EOM_op_XP} analytically with the assumption of fixed point (or steady state) values of the covariance matrix elements and
\begin{equation}
    \boldsymbol{\mathsf{B}}=\begin{pmatrix}\zeta &\Xi\\[4pt]\Xi&\Upsilon \end{pmatrix}
    \label{B_matrix}
\end{equation}for arbitrary measurement strengths gives
\begin{equation}
    \begin{split}
        q_1(\T)&=\Big((\zeta+\Upsilon)\frac{\alpha_1 \T}{2}+q_{1i}\Big)\cos{\T}+\\&\Big((\zeta+\Upsilon)\frac{\alpha_2 \T}{2}+q_{2i}-\big((\Upsilon-\zeta)\frac{\alpha_1}{2}-\Xi\alpha_2\big)\Big)\sin{\T},\\
        q_2(\T)&=\Big((\zeta+\Upsilon)\frac{\alpha_2 \T}{2}+q_{2i}\Big)\cos{\T}- \\&\Big((\zeta+\Upsilon)\frac{\alpha_1 \T}{2}+q_{1i}-\big((\Upsilon-\zeta)\frac{\alpha_2}{2}+\Xi\alpha_1\big)\Big)\sin{\T},\\
        p_1(\T)&=\alpha_1 \cos{\T}+\alpha_2 \sin{\T},\\
        p_2(\T)&=-\alpha_1\sin{\T}+\alpha_2\cos{\T},
    \end{split}
    \label{OP_sol_XP}
\end{equation}where $\alpha_1$ and $\alpha_2$ are integration constants, $q_{1i}$ and $q_{2i}$ are initial values of $q_1$ and $q_2$. Values of $\alpha$'s can be determined from the boundary conditions through the matrix equation\begin{equation}
\begin{split}
    \mathcal{A}(\T_f)\begin{pmatrix}\alpha_1\\ \alpha_2\end{pmatrix} = \begin{pmatrix}q_{1f}-q_{1i}\cos{\T_f}-q_{2i}\sin{\T_f}\\q_{2f}+q_{1i}\sin{\T_f}-q_{2i}\cos{\T_f}\end{pmatrix}.%
   \end{split}
   \label{mat_eqn_XP}%
\end{equation}
\begin{widetext}\hspace{-1em}The elements of the matrix $\mathcal{A}(t_f)$ are
\begin{equation}
    \begin{split}
        &\mathcal{A}_{11}(\T_f)=(\zeta+\Upsilon)\frac{\T_f \cos{\T_f}}{2} -\frac{1}{2}(\Upsilon-\zeta)\sin{\T_f},\quad
         \mathcal{A}_{12}(\T_f)=\Big((\zeta+\Upsilon)\frac{\T_f}{2}+\Xi\Big)\sin{\T_f},\\
         &\mathcal{A}_{21}(\T_f)=\Big(-(\zeta+\Upsilon)\frac{\T_f}{2}+\Xi\Big)\sin{\T_f},\quad \quad\quad\quad ~~
         \mathcal{A}_{22}(\T_f)=(\zeta+\Upsilon)\frac{\T_f \cos{\T_f}}{2} +\frac{1}{2}(\Upsilon-\zeta)\sin{\T_f}.
    \end{split}
    \label{matrix_elements}
\end{equation}
The determinant of $\mathcal{A}(\T_f)$ is 
\begin{equation}
    \det(\mathcal{A}(\T_f))=(\zeta+\Upsilon)^2\frac{\T_f^2}{4}-\Big[\frac{(\Upsilon-\zeta)^2}{4}+\Xi^2\Big]\sin^2{\T_f}.
\label{determinant_XP}
\end{equation}
Comparing the constant term and the coefficient of the sine term we find,
\begin{equation}
    \begin{split}
    \frac{(\zeta+\Upsilon)^2}{4}-\frac{(\Upsilon-\zeta)^2}{4}-\Xi^2=\zeta\Upsilon-\Xi^2
    =\frac{1}{16\tau_1\tau_2}(\tilde{q}_3 \tilde{q}_5-\tilde{q}_4^2)^2=\frac{1}{16\tau_1\tau_2}>0.
    \end{split}
    \label{det_val}
\end{equation}
\\From \eqref{determinant_XP} we find,
\begin{equation}
    \det(\mathcal{A}(\T_f))>\frac{(\zeta+\Upsilon)^2}{4}(\T_f^2-\sin^2{\T_f})\geq 0.
\end{equation}
Hence, the matrix $\mathcal{A}(\T_f)$ is always invertible, giving unique solutions for \eqref{EOM_op_XP}. Consequently, \emph{multipaths\cite{NaghilooCaustic,LewalleMultipath,LewalleChaos} do not exist}.\par
Using \eqref{OP_sol_XP} and \eqref{OP_h_XP}, we can get the stochastic energy (a constant of motion)
\begin{equation}
\begin{split}
    E_s=\frac{1}{2}(\zeta\alpha_1^2+2 \Xi\alpha_1\alpha_2+\Upsilon\alpha_2^2)+(q_{2i}\alpha_1-q_{1i}\alpha_2)-\frac{\tilde{q}_3}{4\tau_1}-\frac{\tilde{q}_5}{4\tau_2}.
\end{split}
    \label{Es}
\end{equation}
The expectation value of the mechanical energy is
\begin{equation}
    E_M(\T)=\frac{1}{4}(\tilde{q}_5+ \tilde{q}_3)+\frac{1}{2}(q_1(\T)^2+q_2(\T)^2).
    \label{e_mech_XP}
\end{equation}

\section{\label{appA4}Probability Density} 
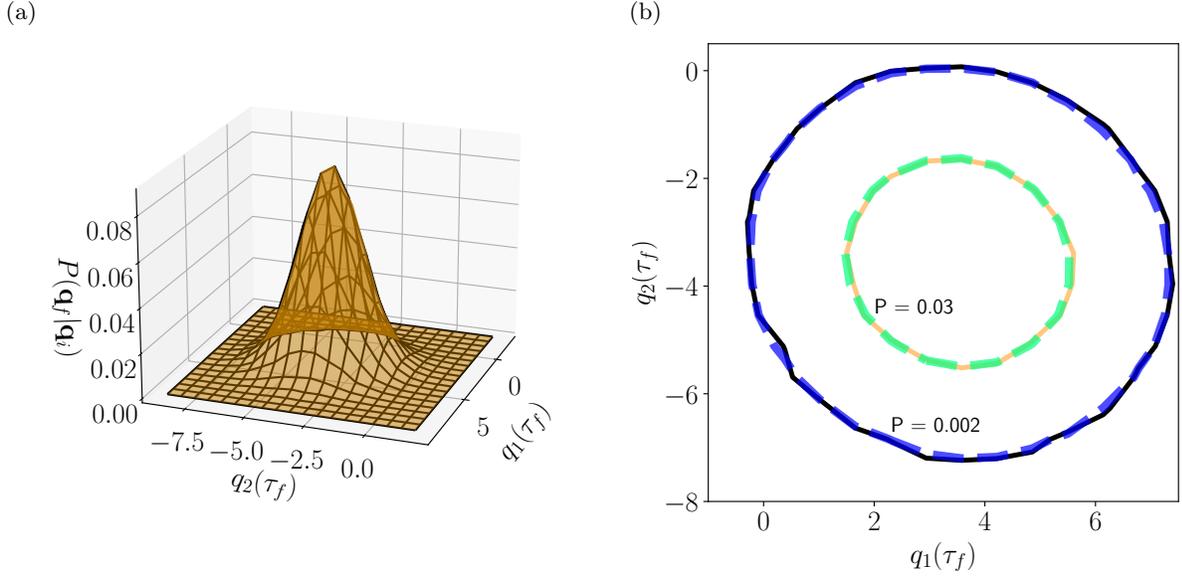
\begin{figure*}
    \includegraphics[width=0.45\textwidth,trim = {0 0 0 50}, clip]{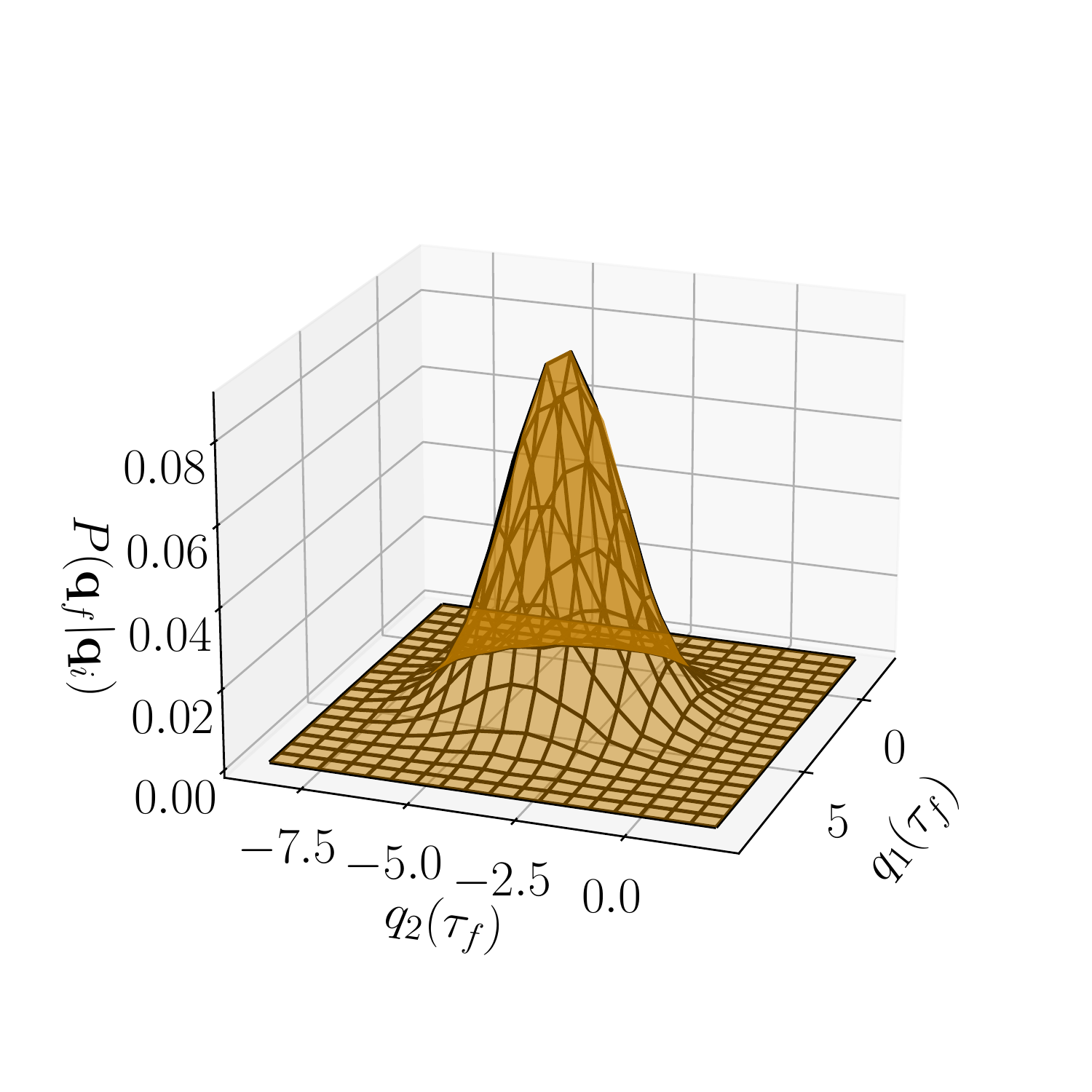}
	\includegraphics[width=0.45\textwidth,trim = {0 0 0 25}, clip]{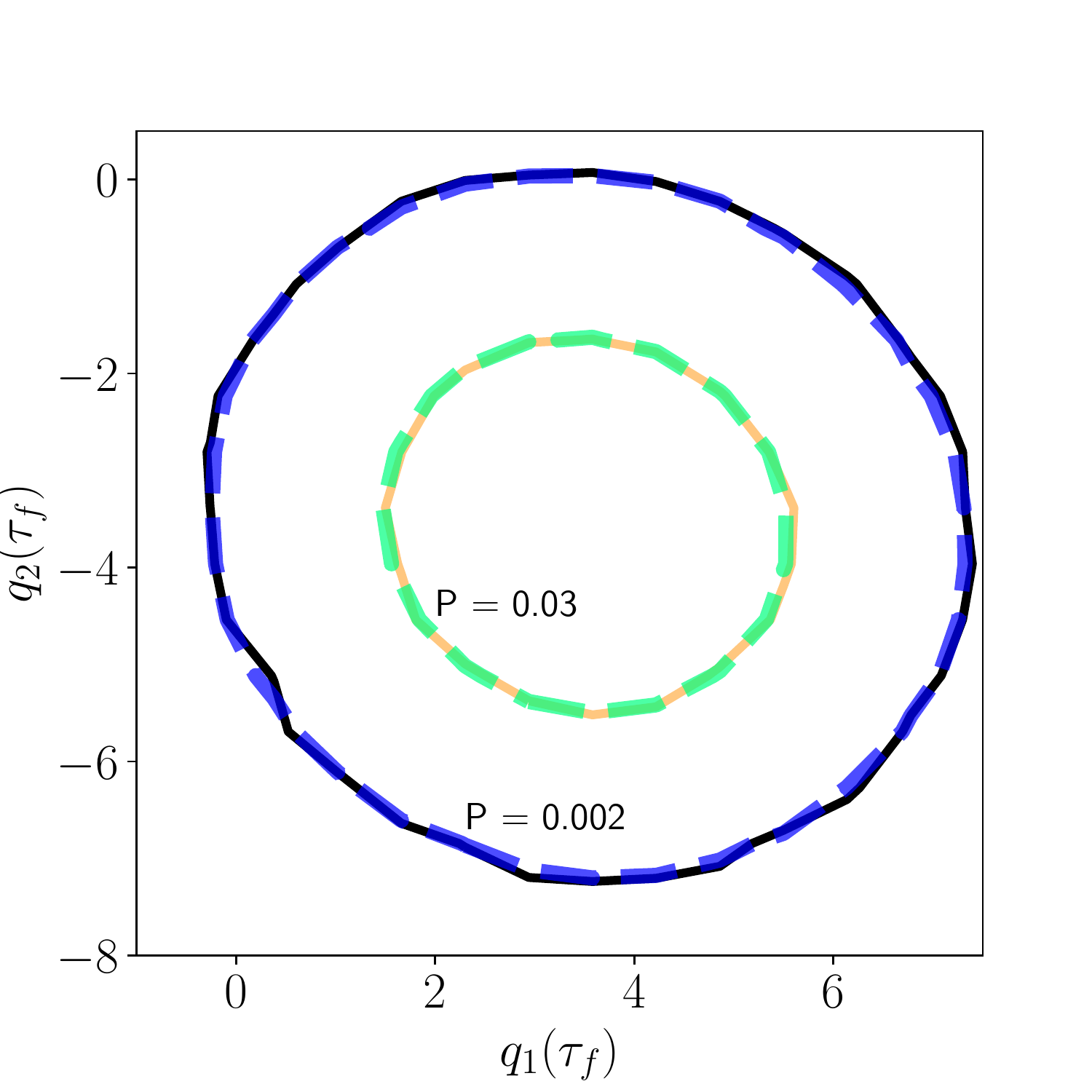}
	\begin{tikzpicture}[overlay]
	\node[] at (-16.3,7.5) {(a)};
	\node[] at (-8,7.5) {(b)};
	\end{tikzpicture}
	\caption{Here we show the probability density of final states at $\T_f=8.0$, starting from the initial state $q_1(0)=3$ and $q_2(0)=4$ for unequal collapse timescales (or equivalently, measurement strengths), arbitrarily chosen to have the values $\T_1=0.7$ and $\T_2=3.0$. Again, in panel (a), a 3-dimensional plot shows the probability density at $\T_f$ in the quadrature space for 100,000 simulated trajectories (black wireframes) and  from the analytical calculations \eqref{P_final_expr} (orange surface). Panel (b) shows the contours for $P(\pmb{q}_f|\pmb{q}_i)=0.03$ and $P(\pmb{q}_f|\pmb{q}_i)=0.002$ for both the simulations (solid lines) and the analytical expression (thick dashed lines). Probability densities calculated from the simulations confirm the validity of our analytical calculations for the general measurement strength case presented in Appendix \ref{appA4}. The difference of the measurement strengths along the two quadratures lead to elliptical or squeezed contour lines, in contrast with the circular contour lines seen in  Fig.~\ref{prob_density_3}.}
	\label{pd_gen}%
\end{figure*}
Here we carry out an explicit calculation of the probability density $P(\pmb{q}_f|\pmb{q}_i)$. From \eqref{pqfqi} and \eqref{stoch_h_XP}, we can write
\begin{equation}
\begin{split}
    P(\pmb{q}_f|\pmb{q}_i)=\int\mathcal{D} \pmb{p}\mathcal{D}\pmb{q}\mathcal{D}\pmb{r}\: \exp\Bigg\{\int_{0}^{\T_f}d\T\Bigg(-\pmb{p}\cdot\pmb{\dot{q}}+\pmb{p}^\top\cdot\left(\boldsymbol{\mathsf{\Omega}}\cdot\pmb{q}+\boldsymbol{\mathsf{b}}\cdot[\boldsymbol{\mathsf{R}} - \boldsymbol{\mathsf{X}}]\right)-\tfrac{1}{2}(\boldsymbol{\mathsf{R}}-\boldsymbol{\mathsf{X}})^2 - \mathsf{g}\Bigg)\Bigg\}.
\end{split}
    \label{p_full_int}
\end{equation}
We note that for the $\pmb{q}$ integral, we choose functions such that $\pmb{q}(0)=\pmb{q}_i$ and $\pmb{q}(\T_f)=\pmb{q}_f$ while no such restrictions apply for the $\pmb{p}$ or $\pmb{r}$ integrals. We expect this integral to have the form 
\begin{equation}
     P(\pmb{q}_f|\pmb{q}_i)=\mathcal{N}f(\pmb{q}_f,\pmb{q}_i,\T_f).
     \label{P_form}
\end{equation}
The normalization constant $\mathcal{N}$ is dependent on the measurement strengths and final time $\T_f$. The functional integral w.r.t. the readouts is Gaussian and can be evaluated to give 
\begin{equation}
\begin{split}
    P(\pmb{q}_f|\pmb{q}_i)=\mathcal{N}_1\int\mathcal{D} \pmb{p}\mathcal{D}\pmb{q}\: \exp\Bigg\{\int_{0}^{\T_f}d\T\Bigg(-\pmb{p}\cdot\dot{\pmb{q}}+\frac{1}{2}\pmb{p}^\top\cdot\boldsymbol{\mathsf{B}}\cdot\pmb{p}+\pmb{p}^\top\cdot\boldsymbol{\mathsf{\Omega}}\cdot \pmb{q}-\mathsf{g}\Bigg)\Bigg\},
    \end{split}
    \label{p_rest_int}
\end{equation} with $\mathcal{N}_1$ as a constant dependent on $\T_f$. Now we expand the quadratures and their conjugate momenta around the OP solution $\pmb{q}=\pmb{q}_{\mathsf{OP}}(\T)+\delta \pmb{q}$ and $\pmb{p}=\pmb{p}_{\mathsf{OP}}(\T)+\delta \pmb{p}$. $\pmb{q}_{\mathsf{OP}}(\T)$ and $\pmb{p}_{\mathsf{OP}}(\T)$ satisfy \eqref{EOM_op_XP} with boundary conditions $\pmb{q}_{\mathsf{OP}}(0)=\pmb{q}_i$, and $\pmb{q}_{\mathsf{OP}}(\T_f)=\pmb{q}_f$. This implies $\delta\pmb{q}(0)=\delta \pmb{q}(\T_f)=0$, with no restrictions on $\delta\pmb{p}$. Using the definitions of $\boldsymbol{\mathsf{B}}$, $\boldsymbol{\mathsf{\Omega}}$, \eqref{EOM_op_XP}, and boundary conditions, \eqref{p_rest_int} can be simplified to 
\begin{equation}
\begin{split}
    P(\pmb{q}_f|\pmb{q}_i)=\mathcal{N}_1\int\mathcal{D} \delta\pmb{p}\mathcal{D}\delta\pmb{q}\: &\exp\Bigg\{\int_{0}^{\T_f}d\T\Bigg(-\delta\pmb{p}\cdot\delta\dot{\pmb{q}}+\frac{1}{2}\delta \pmb{p}^\top\cdot\boldsymbol{\mathsf{B}}\cdot\delta \pmb{p}+\delta \pmb{p}^\top\cdot\boldsymbol{\mathsf{\Omega}}\cdot \delta \pmb{q}\Bigg)\Bigg\}\\&\times\exp\left\{\int_{0}^{\T_f}d\T\Bigg(-\frac{1}{2} \pmb{p}_{\mathsf{OP}}^\top\cdot\boldsymbol{\mathsf{B}}\cdot \pmb{p}_{\mathsf{OP}}-\mathsf{g}\Bigg)\right\}.
    \end{split}
    \label{P_simplified}
\end{equation}
Except for  $\pmb{p}_{\mathsf{OP}}^\top\cdot\boldsymbol{\mathsf{B}}\cdot \pmb{p}_{\mathsf{OP}}$, everything else in the exponential can be absorbed in the normalization constant, giving
\begin{equation}
    P(\pmb{q}_f|\pmb{q}_i)=\mathcal{N}\exp{-\frac{1}{2}\int_0^{\T_f}d\T\: \pmb{p}_{\mathsf{OP}}^\top\cdot\boldsymbol{\mathsf{B}}\cdot \pmb{p}_{\mathsf{OP}}}.
    \label{P_expr1}
\end{equation}
For the steady state case (i.e. fixed point values of $q_3$, $q_4$ and $q_5$), from \eqref{OP_sol_XP} and \eqref{mat_eqn_XP} the OP solution for the momenta are  $\pmb{p}_{\mathsf{OP}}=R(\T)\mathcal{A}(\T_f)^{-1}\Delta \pmb{Q}$ with $R(\T)$ as a rotation matrix and $\Delta \pmb{Q}=(\Delta Q_1,\:\:\Delta Q_2)^\top=(q_{1f}- \mathcal{Q}_{1f},\:\:q_{2f}- \mathcal{Q}_{2f})^\top$. After the time integration in \eqref{P_expr1} and calculating the normalization constant,  the probability density becomes
\begin{equation}
    P(\pmb{q}_f|\pmb{q}_i)=\frac{\sqrt{\det(\Sigma)}}{2\pi}\exp{-\frac{1}{2} \Delta \pmb{Q}^\top\cdot\Sigma\cdot\Delta \pmb{Q}},
    \label{P_final_expr}
\end{equation}
with the matrix

\begin{equation}
    \Sigma=(\mathcal{A}(\T_f)^\top)^{-1}\left\{\frac{\zeta+\Upsilon}{2}\T_f\mathds{1}+\frac{\sin{\T_f}}{2}\begin{pmatrix}(\zeta-\Upsilon)\cos{\T_f}-2\Xi\sin{\T_f} &\:\:\:\:(\zeta-\Upsilon)\sin{\T_f}+2\Xi\cos{\T_f} \\[6pt](\zeta-\Upsilon)\sin{\T_f}+2\Xi\cos{\T_f} &\:\:\:\:-(\zeta-\Upsilon)\cos{\T_f}+2\Xi\sin{\T_f}\end{pmatrix}
    \right\}\mathcal{A}(\T_f)^{-1}.
    \label{sigma_matrix}
\end{equation}
Fig.~\ref{pd_gen} compares our analytical calculations with the simulations of 100,000 trajectories. Our path integral approach correctly predicts the squeezing of the Gaussian probability density function for the general measurement strength case. The probability densities of the individual final quadratures $q_{1f}$ and $q_{2f}$ become
\begin{equation}
     P(q_{1f}|\pmb{q}_i)=\sqrt{\frac{\det(\Sigma)}{2\pi\Sigma_{22}}}\exp{-\frac{\det(\Sigma)}{2\Sigma_{22}}\Delta Q_1^2},\:\:\:\:
     P(q_{2f}|\pmb{q}_i)=\sqrt{\frac{\det(\Sigma)}{2\pi\Sigma_{11}}}\exp{-\frac{\det(\Sigma)}{2\Sigma_{11}}\Delta Q_2^2}.
     \label{P_final_q1q2}
\end{equation}

\section{\label{appA5}Position Measurement}For only position measurement (see Fig.~\ref{fig-device}(a)), $\tau_1=\ct$ and $\tau_2=\infty$. The fixed point values of $q_3$, $q_4$ and $q_5$ are given by 
\begin{equation}
    \begin{split}
        \tilde{q}_3&=\sqrt{4\ct(\sqrt{1+4\ct^2}-2\ct)},\\
        \tilde{q}_4&=\sqrt{1+4\ct^2}-2\ct,\\
        \tilde{q}_5&=\sqrt{\frac{1}{\ct}(\sqrt{1+4\ct^2}-2\ct)(1+4\ct^2)}.
    \end{split}
    \label{fpoints}
\end{equation}The solutions for the optimal path \eqref{OP_sol_XP} are 
\begin{equation}
    \begin{split}
        q_1(\T)&=\Big(\frac{\alpha_1(\tilde{q}_3^2+\tilde{q}_4^2)}{8\ct} \T+q_{1i}\Big)\cos{\T}+\Big(\frac{\alpha_2(\tilde{q}_3^2+\tilde{q}_4^2)}{8\ct}\T+q_{2i}+\frac{\alpha_1(\tilde{q}_3^2-\tilde{q}_4^2)}{8\ct}+\frac{\tilde{q}_3 \tilde{q}_4 \alpha_2}{4\ct}\Big)\sin{\T},\\
        q_2(\T)&=- \Big(\frac{\alpha_1(\tilde{q}_3^2+\tilde{q}_4^2)}{8\ct} \T+q_{1i}+\frac{\alpha_2(\tilde{q}_3^2-\tilde{q}_4^2)}{8\ct}-\frac{\tilde{q}_3 \tilde{q}_4 \alpha_1}{4\ct}\Big)\sin{\T}+\Big(\frac{\alpha_2(\tilde{q}_3^2+\tilde{q}_4^2)}{8\ct}\T+q_{2i}\Big)\cos{\T},\\
        p_1(\T)&=\alpha_1 \cos{\T}+\alpha_2 \sin{\T},\\
        p_2(\T)&=-\alpha_1\sin{\T}+\alpha_2\cos{\T}.
    \end{split}
    \label{OP_sol}
\end{equation}
Just like in the general case, $\alpha_1$ and $\alpha_2$ are integration constants, $q_{1i}$ and $q_{2i}$ are initial values of $q_1$ and $q_2$. Values of $\alpha$'s can be determined from the final conditions through the matrix equation
\begin{equation}
\begin{split}
    \mathcal{A}(\T_f)\begin{pmatrix}\alpha_1\\ \alpha_2\end{pmatrix} = \begin{pmatrix}q_{1f}-q_{1i}\cos{\T_f}-q_{2i}\sin{\T_f}\\q_{2f}+q_{1i}\sin{\T_f}-q_{2i}\cos{\T_f}\end{pmatrix}.
   \end{split}
   \label{mat_eqn}
\end{equation}
The matrix $\mathcal{A}(\T_f)$ is given by
\begin{equation}
\begin{split}
    \mathcal{A}(\T_f)=\frac{1}{8\ct}\begin{pmatrix} (\tilde{q}_3^2+\tilde{q}_4^2)\T_f \cos{\T_f} +(\tilde{q}_3^2-\tilde{q}_4^2)\sin{\T_f} & ((\tilde{q}_3^2+\tilde{q}_4^2)\T_f+2\tilde{q}_3 \tilde{q}_4)\sin{\T_f}\\[8pt] (-(\tilde{q}_3^2+\tilde{q}_4^2)\T_f+2\tilde{q}_3 \tilde{q}_4)\sin{\T_f} & (\tilde{q}_3^2+\tilde{q}_4^2)\T_f \cos{\T_f} -(\tilde{q}_3^2-\tilde{q}_4^2)\sin{\T_f} \end{pmatrix}.
    \end{split}
    \label{matrix}
\end{equation}
The determinant of \eqref{matrix} is 
\begin{equation}
    \det(\mathcal{A}(\T_f))=\frac{(\tilde{q}_3^2+\tilde{q}_4^2)^2}{64\ct^2}(\T_{f}^2-\sin^2{\T_f})>0\:\:\: (\forall \T_f>0),
\label{determinant}
\end{equation}
again proving the non-existence of multipaths.
\end{widetext}
\bibliography{references}
\nocite{*}
\end{document}